\documentclass[12pt,onecolumn,draftclsnofoot]{IEEEtran}

\usepackage{xspace,amsmath,amssymb,amsfonts,epsfig,subfigure,syntonly}
\usepackage{cite,bm,color,url,textcomp,empheq,boxedminipage}
\usepackage{algorithmicx,algorithm}
\usepackage{epstopdf,makecell}
\usepackage{empheq}
\usepackage{pifont}
\usepackage{subeqnarray}
\usepackage{cases}
\usepackage{graphicx,graphics}  
\usepackage{multirow,multicol}
\usepackage{psfrag}    
\usepackage{stfloats}
\usepackage{url}
\usepackage[noend]{algpseudocode}

\begin{document}
\title{\LARGE Design of Link-Selection Strategies for Buffer-Aided DCSK-SWIPT Relay System\vspace{-0.3cm}
\thanks{M.~Qian, G.~Cai, Y.~Fang and G.~Han are with the School of Information Engineering, Guangdong University of Technology, Guangzhou 510006, China (e-mail: mqian95@163.com; caiguofa2006@126.com; fangyi@gdut.edu.cn; gjhan@gdut.edu.cn).}}
\author{\fontsize{11pt}{\baselineskip}\selectfont{Mi Qian, Guofa Cai, Yi Fang, and Guojun Han}}

\date{}
\maketitle
\begin{abstract}
Adaptive link selection for buffer-aided relaying can achieve significant performance gain compared with the conventional relaying with fixed transmission criterion. However, most of the existing link-selection strategies are designed based on perfect channel state information (CSI), which are very complex by requiring channel estimator. To solve this issue, in this paper, we investigate a buffer-aided differential chaos-shift-keying based simultaneous wireless information and power transfer (DCSK-SWIPT) relay system, where a decode-and-forward protocol is considered and the relay is equipped with a data buffer and an energy buffer. In particular, we propose two link-selection protocols for the proposed system based on harvested energy, data-buffer status and energy-shortage status, where the CSI is replaced by the harvested energy to avoid the channel estimation and the practical problem of the decoding cost at the relay is considered. Furthermore, the bit-error-rate (BER) and average-delay closed-form expressions of the proposed protocols are derived over multipath Rayleigh fading channels, which are validated via simulations.
Finally, results demonstrate that both the proposed protocols not only provide better BER performance than the conventional DCSK system and DCSK-SWIPT relay system but also achieve better BER performance and lower average delay in comparison to the conventional signal-to-noise-ratio-based buffer-aided DCSK-SWIPT relay systems.
\end{abstract}


\section{Introduction} \label{sect:review}
Low-cost and low-power short-range wireless cooperative communication has been widely used in the scenarios, such as Internet of Things (IoTs) and wireless body area networks (WBANs) \cite{8782883}. Cooperative communication technology has been recognized as a promising technique because it can enhance the reliability of data transmission and improve the effects of path loss and shadowed fading channels. In general, the cooperative networks depend on either amplify-and-forward (AF) or decode-and-forward (DF) relaying protocol \cite{1362898,7314985}, where the relay receives and transmits the packets in successive time slots. In conventional relaying systems, the bottleneck relay link limits the system performance because the relay nodes employ a fixed protocol for information transmission and reception \cite{8735871}. Hence, to solve the above issue, a data buffer is equipped in the relay to significantly enhance the system performance.

Over the past decade, the buffer-aided relaying systems have been widely studied because it can effectively avoid the unnecessary transmission when the link quality is poor \cite{7765086, 6408173, 8064682, 6613630, 7317590}. A three-node buffer-aided relaying system at the first time has been proposed by properly scheduling the transmissions based on link quality \cite{7765086}, which achieves better performance gain as compared to conventional relaying system.
Moreover, maximum achievable throughput of the buffer-aided relaying system has been analyzed \cite{6408173}, which considers the fixed rate and mixed rate transmission.
In addition, adaptive link-selection protocols for the buffer-aided relaying system have been conceived and their bit error rate (BER) and throughput are optimized \cite{8064682}.
Furthermore, a novel link-selection protocol has been designed for bit-interleaved coded-modulated orthogonal-frequency-division multiplexing buffer-aided relaying system \cite{6613630}, which obtains significant diversity gain.
However, the above buffer-aided relaying systems require perfect channel state information (CSI). It is well known that requiring the CSI at the receiver is generally difficult and complex. To alleviate this issue, two link-selection protocols for buffer-aided relaying system with outdated CSI have been proposed \cite{7317590}.
Recently, a distributed low-complexity link-selection algorithm dealing with outdated CSI for the buffer-aided relaying system has been conceived in \cite{8328875}.

In the buffer-aided relaying systems, the lifetime of a relay node is usually determined by the life cycle of the battery. Since the data stored in the relay node generates energy consumption, the energy of the battery for the relay node is consumed faster than the non-relaying node.
The main sources of traditional energy harvesting (EH) technology are solar and wind energy, but this technology often depends on the weather and is beyond control.
Particularly, in various complicated environments, such as implant WBANs, replacing and charging the battery can be very unrealistic. Therefore, it is necessary to solve the energy-limitation problem of the terminal.
Wireless EH technology has attracted a lot of attention recently as it can provide energy supply for the power-limited devices by using radio frequency (RF) signals \cite{8390912}, which is considered to be a potential candidate because they can simultaneously carry information and energy \cite{8628978,7556958}.
Based on the above advantages, the application of simultaneous wireless information and power transfer (SWIPT) has been studied in many scenarios, such as the two-way relay system \cite{8463536}, the cooperative system \cite{8107546,8060568,8122039} and the broadcasting system \cite{6489506}.
A buffer-aided full-duplex successive relay selection scheme has been proposed and analyzed in energy harvesting IoT networks \cite{8664088}.
In addition, an adaptive buffer-aided wireless powered relay communication with energy storage has been constructed in \cite{8226786}.

However, the majority of the buffer-aided SWIPT relaying systems focus on coherent demodulation and detection techniques. To recover the received information, the coherent systems have to acquire the perfect CSI and realize the carrier synchronization. The corresponding channel estimation and synchronization processes not only result in high complexity of the devices but also need to consume more energy at the relay.
To avoid the channel estimation nor synchronization, the non-coherent technology has been widely used in low-power and low-complexity transmission scenarios.
Recently, many non-coherent modulation technologies, such as frequency-shift-keying \cite{7008567}, differential phase-shift-keying (DPSK) \cite{5896023}, and differential-chaos-shift-keying (DCSK) \cite{8036271} have been widely studied.
The DCSK system has attracted a great deal of attention due to the advantages of low complexity and low cost, which is the same as the DPSK system.
In particular, the DCSK system is more robust to multipath fading channel as compare with the DPSK system \cite{6482240,1362943,7426400,5629387}, thus having been considered as an excellent alternative modulation for ultra-wideband systems \cite{7442517}. Thanks to the above advantages, many DCSK variants have been proposed in \cite{8606201,8703732,8110728,8468068,8618388,7973179,7604059,7109922}.

To substantially exploit the benefits of DCSK modulation and buffer-aided SWIPT relaying system, this paper investigates a buffer-aided DCSK-SWIPT relay system, which can achieve excellent BER performance.
Moreover, there are many literatures studying the link-selection strategies based on the signal-to-noise-ratio (SNR) information \cite{6807959,7365416,8830394}. To avoid the channel estimation at the receiver, this paper proposes two link-selection protocols without using any CSI. The contributions of this paper are summarized as follow:

\begin{enumerate}
\item
A buffer-aided DCSK-SWIPT relay system is put forward, where the DF relay has functionalities of data buffer and energy buffer. To make the system more practical, the decoding cost is considered at the relay, which may lead to the energy shortage at the scenario of the harvested energy that cannot satisfy the energy cost by the relay decoder (i.e., decoding cost). Furthermore, two link-selection strategies of the proposed system without using CSI are conceived based on harvested energy, data-buffer status and energy-shortage status, where harvested energy is used instead of CSI.

\item
Closed-form BER expressions of the two link-selection strategies for the proposed system are derived over multipath Rayleigh fading channels by using Meijer G-function and Guass-Hermite approach. Moreover, the closed-form expressions of the average delay considering both the queuing delay and the silent time slots are derived.

\item
Simulations are carried out to verify the accuracy of the theoretical analysis and demonstrate that the proposed two protocols possess better BER performance in comparison to the DCSK system and conventional DCSK-SWIPT relay system. Furthermore, the proposed schemes not only realize better BER, but also obtain lower average delay compared to the conventional SNR-based ones. In addition, the impact of the underlying system parameters on BER and average-delay performance is discussed and some important insights are provided.

\end{enumerate}
%
%

The remainder of this paper is organized as follows. A buffer-aided DCSK-SWIPT relay system and two link-selection protocols are presented in Section II. BER and average-delay performance analysis are reported in Section III and Section IV, respectively. Simulation results and discussions are summarized in Section V. Finally, the conclusion is provided in Section VI.

\begin{figure}[tbp]
\center
\includegraphics[width=2.7in,height=1.8in]{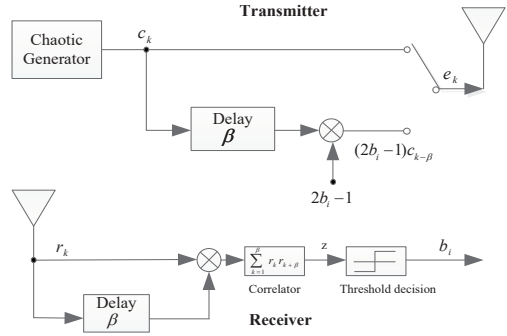}
\caption{Block diagram of the DCSK transceiver}
\label{fig:fig1}
\vspace{-0.5cm}
\end{figure}

\section{System Model} \label{sect:system model}
In this section, the basic principle of the DCSK system is first introduced. Then, the signal model of the proposed buffer-aided DCSK-SWIPT relay system is proposed. Finally, two new link-selection strategies are conceived.\vspace{-1.5mm}

\subsection{DCSK System} \label{sect:conv_DCSK}\vspace{-0.5mm}
In DCSK modulation, the $i$-th transmitted bit, i.e., $b_{i}\in \left \{ 0,1 \right \}$, is represented by two sets of chaotic sequences. Specifically, in the first half bit duration, the transmitter sends a reference sequence; in the second half bit duration, if bit '0' is transmitted, the information sequences equal to the reference sequences, otherwise an inverted version of reference sequences is delivered. In this paper, logistic map, i.e., $c_{k+1}=1-2c_{k}^{2}$, is employed to generate a chaotic signal, where $c_k ( k=1,2,...)$ denotes the $k$-th chip of a chaotic signal. Let $2\beta $ be the spreading factor, the $k$-th chip of the transmitted signal for the DCSK system can be written by
\begin{equation}
e_{k}=\left\{
\begin{aligned}
&\hspace{1mm}c_{k},~~~~~ \hspace{-0.4mm}k=2(i-1)\beta+1,\ldots, 2(i-1)\beta+\beta\\
&\left(2b_i-1\right)c_{k-\beta},~~~k=(2i-1)\beta+1,\ldots, 2i\beta
\end{aligned}
\right..
\label{eq:1func}
\end{equation}
Then, the transmitted signal is passed through a channel, and the received signal $r_k$ is obtained. Furthermore, $r_k$ is correlated with its delayed version and the correlated results are summed over a half bit duration. Hence, one can get a decision metric, given by
\begin{equation}
z=\sum_{k=1}^{\beta}r_{k}r_{k+\beta}.
 \label{eq:2func}
\end{equation}
Finally, the estimated information is obtained by using (\ref{eq:2func}).
\begin{figure*}[tbp]
\center
\includegraphics[width=5.2in,height=2.3in]{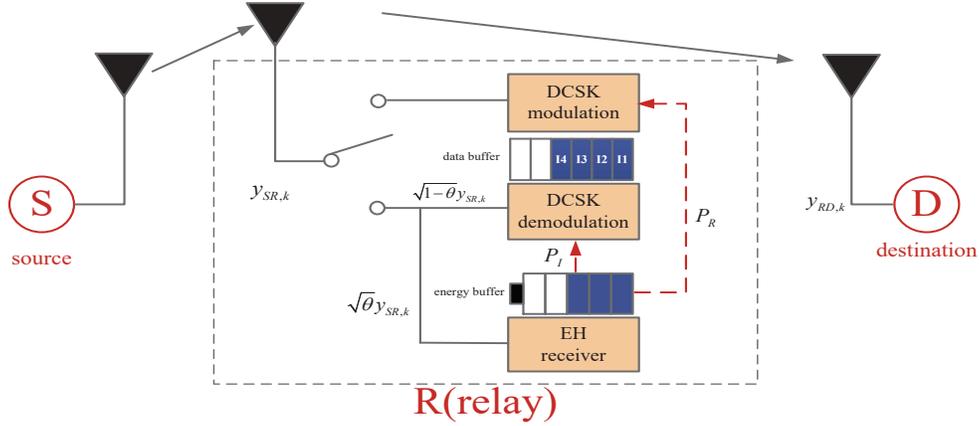}
\vspace{-0.2cm}
\caption{Buffer-aided DCSK-SWIPT relay system model.}
\label{fig:Fig2}\vspace{-0.6cm}
\end{figure*}

\subsection{Buffer-Aided DCSK-SWIPT Relay System}
We consider a buffer-aided DCSK-SWIPT relay system as illustrated in Fig.~\ref{fig:fig1}, which consists of a source S, a relay R, and a destination D. In this system, direct link between S and D is not exploited due to heavy attenuation and all nodes are equipped with single-antenna DCSK system.
Moreover, it is assumed that both S and D have fixed power supplies while R is an energy-constrained node, thus R must be equipped with an energy buffer for storing the harvested energy from the RF signal of S.
Furthermore, R has a data buffer of finite size $J$ for storing data packets, i.e., the buffer has $J$ elements and each element can store one packet of information bits, thus it implies that R must adopt harvest-store-use (HSU) structure to harvest, store and use the energy of the battery \cite{8718530}.
In practical applications a sensor can be used to accurately measure the amount of energy in the battery at each time slot. That is to say, at each time slot the amount of the harvested energy can be recoded. Hence, the status of the data buffer can correspond with that of the energy buffer.
In addition, to simplify the analysis, we assume that S always has data to transmit.
In this system, the proposed link-selection protocols in Sect.~II-B determine whether R receives a data packet from S or sends a data packet to D. Hence, the signal model of the proposed system is in detail depicted as follows. In each time slot, the wireless fading channels are unchanged and have reciprocity.

If the S$\rightarrow$R link is selected at the time slot $q$, S transmits the DCSK-modulated signal to R. The transmitted signal, i.e., $e_{SR,k}$, is obtained by using (\ref{eq:1func}), where $b_i$ is instead of the transmitted bit.
The received signal at R can be expressed as
\begin{equation}
y_{SR,k} = \sqrt {\frac{{{P_S}}}{{d_{sr}^\alpha }}}  \sum\limits_{l  = 1}^{L_{sr}} {h_{sr,l }} e_{SR, k - {\tau _{sr,l }}} + n_{SR,k} ,
\label{eq:3func}
\end{equation}
where $P_S$ denotes the transmitted power at S, $d_{sr}$ is the distance between S and R, $\alpha$ is the pass loss coefficient, $L_{sr}$ is the number of path for channel, $\tau _{sr,l }$ and $h_{sr,l }$ denote delay and channel fading coefficients of the $l$-th path for S$\rightarrow$R link, respectively, and $n_{SR, k}$ represents the additive white Gaussian noise (AWGN) with zero mean and variance $\frac{N_{0, sr}}{2}$.

Then, the received signal at R is divided into two parts: one is forwarded to the energy harvester and the other is adopted for decoding information.
For the former, the available power at EH receiver is calculated as
\begin{equation}
{P_{SR, EH}} = \frac{{\eta \theta {P_S}\sum\limits_{l  = 1}^{L_{sr}} {{h_{sr,l }^2}} }}{{d_{sr}^\alpha }},
\label{eq:4func}
\end{equation}
where $\eta \left ( 0 \le \eta  \le 1 \right )$ denotes the energy conversion efficient factor.
For the latter, the received signal at the information-decoding (ID) receiver~(i.e., DCSK demodulation) of R is formulated as
\begin{equation}
\begin{aligned}
y_{IR,k} &=\sqrt{1-\theta }y_{SR,k} +n_{SI,k}
=\sqrt{\frac{\left ( 1-\theta  \right )P_{S}}{d_{sr}^{\alpha }}}\sum_{l=1}^{L_{sr}}h_{sr,l}
e_{SR,{k-\tau_{sr,l}}} +n_{IR,k},
\label{eq:5func}
\end{aligned}
\end{equation}
where $\theta \left ( 0 \le \theta \le 1 \right )$ denotes the power-splitting ratio, $n_{SI,k}$ is the AWGN with zero mean and variance $\frac{N_{0,SI}}{2}$ due to the RF to baseband signal conversion, $n_{IR,k}=\sqrt{1-\theta }n_{SR,k} + n_{SI,k}$ presents the overall noise of R with zero mean and variance $\frac{N_{0,IR}}{2}$, and $N_{0,IR}=\left ( 1-\theta  \right )N_{0,sr}+N_{0,SI}$.
In fact, to demodulate the received signal $y_{IR,k}$, an energy should be consumed, which is defined as $P_I$. Hence, the remaining energy is given by  $P_R = P_{SR, EH}-P_I$, which is used for the transmitted power at R.

\begin{table*}[!ht]
\center
\caption{Link-Selection Protocol 1. $X_1$ means the link selection is not related to the energy judge. $X_2$ means the link selection is not related to the energy shortage. $X_3$ means the link selection is not related to both energy shortage and energy judge. $X_4$ represents a silent time slot. A means that one packet is added in the data buffer. R means that the data buffer is reduced by one packet.}\vspace{-1mm}
\label{table1}

\begin{tabular}{|c|c|c|c|c|}
\hline
\rule{0pt}{10pt}
Case & Buffer status & Energy judge \& Energy shortage & Link selected & Data buffer \\ \hline
\rule{0pt}{10pt}
\multirow{2}{*}{1} & \multirow{2}{*}{$\phi _j=0$} & $X_1~\&~ P_{SR,EH}> P_I$ & S$\rightarrow$R & A \\ \cline{3-5}
\rule{0pt}{10pt}
 &  & $X_1~\&~ P_{SR,EH}< P_I$ & \multicolumn{2}{c|}{$X_4$} \\ \hline
 \rule{0pt}{10pt}
2 & $\phi _j=J$ & $X_3$ & R$\rightarrow$D & R \\ \hline
\rule{0pt}{10pt}
\multirow{3}{*}{3} & \multirow{3}{*}{$\phi _j\neq \left\{0,J\right\}$} & $P_{SR,EH}\geq \delta  P_{DR,EH}~\&~ P_{SR,EH}> P_I$ & S$\rightarrow$R & A \\ \cline{3-5}
\rule{0pt}{10pt}
 &  & $P_{SR,EH}\geq \delta  P_{DR,EH}~\&~ P_{SR,EH}< P_I$ & \multicolumn{2}{c|}{$X_4$} \\ \cline{3-5}
 \rule{0pt}{10pt}
 &  & $P_{SR,EH}<\delta  P_{DR,EH}~\&~ X_2$ & R$\rightarrow$D & R \\ \hline
\end{tabular}\vspace{-0.3cm}
\end{table*}

\begin{table*}[!htbp]
\center
\caption{Link-selection Protocol 2. $X_2$, $X_4$, A and R are the same as the definitions in TABLE I.}\vspace{-1mm}
\label{tablell}

\begin{tabular}{|c|c|c|c|c|}
\hline
\rule{0pt}{10pt}
Case & Buffer status & Energy judge \& Energy shortage & Link selected & Data buffer \\ \hline
\rule{0pt}{10pt}
\multirow{3}{*}{1} & \multirow{3}{*}{$\phi _j=0$} & $P_{SR,EH}\geq \delta  P_{DR,EH}~\&~ P_{SR,EH}>  P_I$ & S$\rightarrow$R & A \\ \cline{3-5}
\rule{0pt}{10pt}
 &  & $P_{SR,EH}\geq \delta  P_{DR,EH}~\&~ P_{SR,EH}<  P_I$ & \multicolumn{2}{c|}{\multirow{4}{*}{$X_4$}} \\ \cline{3-3}
 \rule{0pt}{10pt}
 &  & $P_{SR,EH}<\delta  P_{DR,EH}~\&~ X_2$ & \multicolumn{2}{c|}{} \\ \cline{1-3}
 \rule{0pt}{10pt}
\multirow{3}{*}{2} & \multirow{3}{*}{$\phi _j=J$} & $P_{SR,EH}\geq \delta  P_{DR,EH} ~\&~ P_{SR,EH}>  P_I$ & \multicolumn{2}{c|}{} \\ \cline{3-3}
\rule{0pt}{10pt}
 &  & $P_{SR,EH}\geq \delta  P_{DR,EH}~\&~ P_{SR,EH}<  P_I$ & \multicolumn{2}{c|}{} \\ \cline{3-5}
 \rule{0pt}{10pt}
 &  & $P_{SR,EH}<\delta  P_{DR,EH}~\&~ X_2$ & R$\rightarrow$D & R \\ \hline
 \rule{0pt}{10pt}
\multirow{3}{*}{3} & \multirow{3}{*}{$\phi _j\neq \left\{0,J\right\}$} & $P_{SR,EH}\geq \delta  P_{DR,EH}~\&~ P_{SR,EH}>  P_I$ & S$\rightarrow$R & A \\ \cline{3-5}
\rule{0pt}{10pt}
 &  & $P_{SR,EH}\geq \delta  P_{DR,EH}~\&~ P_{SR,EH}<  P_I$ & \multicolumn{2}{c|}{$X_4$} \\ \cline{3-5}
 \rule{0pt}{10pt}
 &  & $P_{SR,EH}<\delta  P_{DR,EH}~\&~ X_2$ & R$\rightarrow$D & R \\ \hline
\end{tabular}
\end{table*}

If the R$\rightarrow$D link is selected at the time slot $p (p>q)$, the information transmitter forwards the remodulated signal to D. The remodulated signal, i.e., $e_{RD, k}$, can be obtained by utilizing (\ref{eq:1func}), where $b_i$ comes from the estimated bits stored in the buffer of R. Therefore, the signal received at D can be written as
\begin{equation}
{y_{RD,k}} = \sqrt {\frac{{{P_R}}}{{d_{rd}^\alpha }}} \sum\limits_{l  = 1}^{L_{rd}} {h_{rd,l }} e_{RD,k- \tau _{rd,l}}  + n_{RD,k},
\label{eq:6func}
\end{equation}
where $d_{rd}$ is the distance between R and D, $L_{rd}$ is the number of path for channel, $\tau _{rd,l }$ and $h_{rd,l }$ denote delay and channel fading coefficients of the $l$-th path for R$\rightarrow$D link, respectively, and $n_{RD, k}$ represents the AWGN with zero mean and variance $\frac{N_{0, rd}}{2}$.

\subsection{Two Link-Selection Criterions} \label{sect:DCSK-CC}
For conventional buffer-aided relay system, the instantaneous SNR of S$\rightarrow$R and R$\rightarrow$D links, i.e., $\gamma_{SR}$ and $\gamma_{RD}$ given in Sect.~III, which can be computed by utilizing (\ref{eq:5func}) and (\ref{eq:6func}), respectively, are used for reliability metric. In other word, the conventional system needs to obtain the CSI through performing channel estimation.
However, due to the non-coherent feature for the proposed system and considering the decoding cost, it is difficult to obtain the CSI.
It can be seen from Eq.~(\ref{eq:4func}) that the harvested energy implicitly contains the information of wireless channel. To avoid the channel estimation, harvested energy is used for reliability metric instead of the SNR.
To obtain the harvested energy, in this paper R is employed to collect the information of harvested energy and feeds the result of the link-selection process back to S and D.
Specifically, in the phase~$1$, S transmits a pilot signal to R and the harvested energy $P_{SR,EH}$, i.e., Eq.~(\ref{eq:4func}), of S$\rightarrow$R link can be measured at R. In the phase~$2$, D transmits a pilot signal to R and the harvested energy $P_{DR,EH}$ of R$\rightarrow$D link can be measured by using the reciprocity of wireless channels at R, is given by
\begin{equation}
{P_{DR, EH}} = \frac{{\eta \theta {P_D}\sum\limits_{l  = 1}^{L_{rd}} {{h_{rd,l }^2}} }}{{d_{rd}^\alpha }},
\label{eq:7func}
\end{equation}
where $P_D$ is the transmitted power and $P_D = P_S$.
Through the above operations, R can obtain reliability metrics of the S$\rightarrow$R and R$\rightarrow$D links by using the harvested energy. In the phase~3, R obtains the link-selection result based on the harvested energy, the decoding cost and the data-buffer status, and then R informs the result to S and D, where the transmitted information for the result is powered by using the harvested energy.

Differing from conventional link-selection criterions, the proposed link-selection criterions are based on harvested energy, data-buffer status and energy-shortage status, where energy-shortage status denotes the case of $P_{SR,EH} \textless P_I$.

\subsubsection{Protocol 1} \label{sect:DCSK-ARQ/CARQ}
As depicted in Table I, the information transmission from S to D is divided into three cases, where $\phi _j$ denotes the number of buffer elements at $j$-th time slot.
For $0 < \phi _{j}< J$, the S$\rightarrow$R link is selected when $P_{SR,EH}> \delta  P_{DR,EH} ~\&~ P_{SR,EH}>P_I$, and R$\rightarrow$D link is selected just when $P_{SR,EH}< \delta  P_{DR,EH}$ holds, where $ \delta $ is a decision threshold that can be adjusted to balance the selection of both links and $a~\&~b$ denotes that $a$ and $b$ are simultaneously true. However, when $P_{SR,EH}> \delta  P_{DR,EH} ~\&~ P_{SR,EH}<P_I$, the outage of S$\rightarrow$R link occurs because the harvested energy is insufficient to recover the received signal. For $\phi _{j}=0$, the link selection is not correlated with harvested energy, i.e., the S$\rightarrow$R link is selected just when $P_{SR,EH}>P_I$ holds. Similarly, the R$\rightarrow$D link is selected when $\phi _{j} = J$.

\subsubsection{Protocol 2} \label{sect:DCSK-ARQ}
Unlike the Protocol~$1$, we do not force S and R to transmit information in case of poor link quality. Hence, Protocol~2 has lower BER at the cost of average delay, which is discussed in Sect.~IV. As depicted in Table~II, the information transmission from S to D is also divided into three cases. For $0 < \phi _{j}< J$, the S$\rightarrow$R link is selected only when $P_{SR,EH}> \delta  P_{DR,EH} ~\&~ P_{SR,EH}>P_I$, and the R$\rightarrow$D link is selected just when $P_{SR,EH}< \delta P_{DR,EH}$ holds. For $\phi _j=0$, the S$\rightarrow$R link is selected only when $P_{SR,EH}> \delta   P_{DR,EH} ~\&~ P_{SR,EH}>P_I$ holds. For $\phi _j=J$, the R$\rightarrow$D link is selected if $P_{SR,EH}< \delta   P_{DR,EH}$. For all other cases, the system is in a static mode because no information is transmitted.

\section{BER Analysis} \label{sect:error rate}
In this section, the closed-form BER expressions of the buffer-aided DCSK-SWIPT relay system are derived over multipath Rayleigh fading channels. We assume that the largest delay of the channel is much shorter than the symbol duration, i.e., $0<\tau_{L_{\epsilon}, max}\ll \beta$, $\epsilon \in \{sr, rd\}$, and thus the inter-symbol interference (ISI) can be negligible.
Without loss of generality, it is also assumed that the channel gains of all links are equal, i.e., $E\left \{ h _{1}^{2} \right \}= \ldots =E\left \{ h _{L_{\epsilon}}^{2} \right \}$. It should be noted that the same derivation method can be used for the other channel conditions.

\subsection{Energy Shortage Probability }
According to (\ref{eq:4func}), the outage of S$\rightarrow$R link occurs when the harvested energy is insufficient to recover the received signal. Hence, the energy shortage probability expression is calculated as
\begin{equation}
{P_{ES}} = \Pr \left\{ {{P_{SR,EH}} < {P_I}} \right\} = \Pr \left\{ {\frac{{\eta \theta {P_S}\sum\limits_{l  = 1}^{{L_{sr}}} {{h_{sr,l }^2}} }}{{d_{sr}^\alpha }} < {P_I}} \right\},
\label{eq:8func}
\end{equation}
where $\Pr\{\varphi\}$ denotes the probability of $\varphi$.

Because $\sum_{l=0}^{L_{sr}}h_{sr,l}^{2}$ follows the chi-square distribution with $2L_{sr}$ degrees of freedom, i.e., $\sum_{l=0}^{L_{sr}}h_{sr,l}^{2}\sim \chi_{2L_{sr}}^{2}$, through simple calculation, the closed-form expression of Eq.~\eqref{eq:8func} is computed as
\begin{equation}
{P_{ES}} = \frac{{\gamma \left( {L_{sr},\frac{{{P_I}L_{sr}d_{sr}^\alpha }}{{\eta \theta {P_S}}}} \right)}}{{\Gamma \left( L_{sr} \right)}},
\label{eq:9func}
\end{equation}
where $\gamma \left ( \mu,K \right )=\int_{0}^{K}u^{\mu-1}e^{-u}du$ and $\Gamma \left ( \mu \right )=\int_{0}^{\infty }u^{\mu-1}e^{-\mu}du$ denote the lower incomplete Gamma function and the Gamma function, respectively.\vspace{-3mm}

\subsection{BER Derivation of the DCSK-SWIPT System}
According to \eqref{eq:2func} and \eqref{eq:5func}, the output of the correlator at R can be formulated as
\begin{equation}
\vspace{1mm}
\begin{aligned}
&\hspace{1mm}{Z_r} = \sum\limits_{k = 1}^\beta  {{y_{IR,k}} * {y_{IR,k + \beta }}} \\
 &= \sum\limits_{k = 1}^\beta  {\left\{ {\frac{{\left( {1 - \theta } \right){P_S}}}{{d_{sr}^\alpha }}\sum\limits_{l  = 1}^{{L_{sr}}} {h_{_{sr,l }}^2} {e_{SR,k - {\tau _{sr,l }}}}{e_{SR,k + \beta  - {\tau _{sr,l }}}}} \right.} \\
& + \sqrt {\frac{{\left( {1 - \theta } \right){P_S}}}{{d_{sr}^\alpha }}} \sum\limits_{l  = 1}^{{L_{sr}}} {{h_{sr,l }}} {e_{SR,k - {\tau _{sr,l }}}}{n_{IR,k + \beta }} + {n_{IR,k}}{n_{IR,k + \beta }}\\
&\left. { + \sqrt {\frac{{\left( {1 - \theta } \right){P_S}}}{{d_{sr}^\alpha }}} \sum\limits_{l  = 1}^{{L_{sr}}} {{h_{sr,l }}} {e_{SR,k + \beta  - {\tau _{sr,l }}}}{n_{IR,k}}} \right\}.
\label{eq:10func}
\end{aligned}
\end{equation}

Based on (\ref{eq:10func}), the mean and the variance of $Z_r$ are given by
\begin{equation}
\begin{aligned}
E\left\{ {{Z_r}} \right\} &= \frac{{\left( {1 - \theta } \right){P_S}}}{{d_{sr}^\alpha }}\sum\limits_{k = 1}^\beta  {\sum\limits_{l  = 1}^{{L_{sr}}} {h_{_{sr,l }}^2E\left\{ {{e_{SR,k - {\tau _{sr,l }}}}{e_{SR,k + \beta  - {\tau _{sr,l }}}}} \right\}} } \\
 &= \frac{{\left( {1 - \theta } \right){P_S}}}{{2d_{sr}^\alpha }}\sum\limits_{l  = 1}^{{L_{sr}}} {h_{_{sr,l }}^2},
\label{eq:11func}
\end{aligned}
\end{equation}

\begin{equation}
\begin{aligned}
Var\left\{ {{Z_r}} \right\} &= \sum\limits_{k = 1}^\beta  {\left\{ \begin{array}{l}
\frac{{\left( {1 - \theta } \right){P_S}{N_{0,IR}}}}{{2d_{sr}^\alpha }}\sum\limits_{l  = 1}^{{L_{sr}}} {h_{_{sr,l }}^2} E\left\{ {e_{_{SR,k - {\tau _{sr,l }}}}^2} \right\}\\
 + \frac{{\left( {1 - \theta } \right){P_S}{N_{0,IR}}}}{{2d_{sr}^\alpha }}\sum\limits_{l  = 1}^{{L_{sr}}} {h_{_{sr,l }}^2} E\left\{ {e_{_{SR,k + \beta  - {\tau _{sr,l }}}}^2} \right\}\\
 + \frac{{N_{_{0,IR}}^2}}{4}
\end{array} \right\}} \\
 &= \frac{{\left( {1 - \theta } \right){P_S}\sum\limits_{l  = 1}^{{L_{sr}}} {h_{_{sr,l }}^2} {N_{0,IR}}}}{{2d_{sr}^\alpha }} + \frac{{\beta N_{_{0,IR}}^2}}{4},
 \label{eq:12func}
\end{aligned}
\end{equation}
where $E \{\cdot\}$ and $Var\{ \cdot \}$ denote the expectation and variance operators, respectively. Hence, the BER of the information decoding on S$\rightarrow$R link can be obtained as

\begin{equation}
\begin{aligned}
&{P_{e, SR}^{IR}} = \frac{1}{2}erfc\left( {{{\left[ {\frac{{2Var\left\{ {{Z_r}} \right\}}}{{{E^2}\left\{ {{Z_r}} \right\}}}} \right]}^{ - \frac{1}{2}}}} \right)\\
 &= \frac{1}{2} erfc \left( {{{\left[ {\frac{8}{{\frac{{2\left( {1 - \theta } \right){P_S}\sum\limits_{l  = 1}^{{L_{sr}}} {{h_{sr,l }^2}} }}{{d_{sr}^\alpha {N_{0,IR}}}}}} + \frac{{8\beta }}{{{{\left( {\frac{{2\left( {1 - \theta } \right){P_S}\sum\limits_{l  = 1}^{{L_{sr}}} {{h_{sr,l }^2}} }}{{d_{sr}^\alpha {N_{0,IR}}}}} \right)}^2}}}} \right]}^{ - \frac{1}{2}}}} \right)\\
 &= \frac{1}{2}erfc\left( {{ {\left[ {\frac{8}{{{\gamma _{SR}}}} + \frac{{8\beta }}{{\gamma _{SR}^2}}} \right]}^{ - \frac{1}{2}}}} \right),
 \label{eq:13func}
\end{aligned}
\end{equation}
where $erfc(x)=\frac{2}{\sqrt{\pi }}\int_{x}^{\infty }e^{-t^{2}}dt$, for $x \geq 0$, and
$\gamma_{SR}$ denotes the instantaneous SNR of S$\rightarrow$R link, written as $
{\gamma _{SR}} = \frac{{2\left( {1 - \theta } \right){P_S}\sum\limits_{l  = 1}^{{L_{sr}}} {h_{sr,l }^2} }}{{d_{sr}^\alpha {N_{0,IR}}}}$.

Similarly, the BER of R$\rightarrow$D link is formulated as
\begin{equation}
{P_{e, RD}} = \frac{1}{2}erfc\left( {{{\left[ {\frac{8}{{{\gamma _{RD}}}} + \frac{{8\beta }}{{\gamma _{RD}^2}}} \right]}^{ - \frac{1}{2}}}} \right),
\label{eq:14func}
\end{equation}
where $\gamma_{RD}$ denotes the instantaneous SNR of R$\rightarrow$D link, given by ${\gamma _{RD}} = \frac{{2{P_R}\sum\limits_{l  = 1}^{{L_{rd}}} {h_{rd,l}^2} }}{{d_{rd}^\alpha {N_{0,rd}}}}$.

\subsection{BER of the Protocol~$1$} \label{sect:Performance}
For the S$\rightarrow$R link, the decoding error occurs only when the harvested energy is sufficient to recover the received signal. Hence, the BER expression of S$\rightarrow$R link can be given by
\begin{equation}
{P_{e, SR}} = \left( {1 - {P_{ES}}} \right){P_{e, SR}^{IR}}.
\label{eq:15func}
\end{equation}

According to the status of the data-buffer, the transmission of a source packet from S to D is divided into four cases. The BER of different cases is defined as $P_{f}$, where $f\in \left \{ 1,2,3,4 \right \}$. It should be noted that $P_f$ depends on the energy shortage probability and the probabilities of the buffer being full and empty for the Protocol~$1$, which is defined as $P_{full,P1}$ and $P_{empty,P1}$, respectively. And the closed-form expressions of $P_{full,P1}$ and $P_{empty,P1}$ are derived in Appendix A.
The S$\rightarrow$R and R$\rightarrow$D links are selected according to Table I.

\textbf{Case}~1: When the buffer is neither full nor empty, $P_1$ can be obtained as
\begin{equation}
{P_1} = \left( {1 - {P_{empty,P1}}} \right)\left( {1 - {P_{full,P1}}} \right)\left( {P_{e, SR}^{'} + P_{e, RD}^{'}} \right).
\label{eq:16func}
\end{equation}
In this case, link selection not only depends on harvested energy, but also on the status of the data-buffer. Based on the link-selection protocol for $0< \phi _{j}< J$, $P_{e, SR}^{'}$ and $P_{e, RD}^{'}$ is respectively given by

\begin{equation}
\begin{aligned}
\hspace{15mm}P_{_{e,SR}}^{'} &\le E\left\{ {\left[ {\frac{{\left( {1 - {P_{ES}}} \right)}}{2}erfc\left( {{{\left[ {\frac{8}{{{\gamma _{SR}}}} + \frac{{8\beta }}{{\gamma _{SR}^2}}}\right]}^{ - \frac{1}{2}}}} \right)} \right]|{P_{SR,EH}} > \delta {P_{DR,EH}}} \right\}\\
 &= \frac{{{K_1}\overline P _{SR,EH}^{{L_{rd}}}}}{{\Gamma \left( {{L_{rd}}} \right)}}\sum\limits_{l = 0}^{{L_{sr}} - 1} {\frac{{{\delta ^l}\overline P _{DR,EH}^{l}\Gamma \left( {{L_{rd}} + l} \right)}}{{l!{{\left( {\delta {{\overline P }_{DR,EH}} + {{\overline P }_{SR,EH}}} \right)}^{l + {L_{rd}}}}}}},
\label{eq:17func}
\end{aligned}
\end{equation}

\begin{equation}
\begin{aligned}
\hspace{-2mm}P_{_{e,RD}}^{'} &\le E\left\{ {\frac{1}{2}erfc\left( {{{\left[ {\frac{8}{{{\gamma _{RD}}}} + \frac{{8\beta }}{{\gamma _{RD}^2}}} \right]}^{ - \frac{1}{2}}}} \right)|{P_{DR,EH}} > {P_{SR,EH}}/\delta } \right\}\\
 &= \frac{{{K_2}{\delta ^{{L_{sr}}}}\overline P _{DR,EH}^{{L_{sr}}}}}{{\Gamma \left( {{L_{sr}}} \right)}}\sum\limits_{l = 0}^{{L_{rd}} - 1} {\frac{{\Gamma \left( {{L_{sr}} + l} \right)\overline P _{SR,EH}^{l}}}{{l!{{\left( {{{\overline P }_{SR,EH}} + \delta {{\overline P }_{DR,EH}}} \right)}^{l + {L_{sr}}}}}}},
 \label{eq:18func}
\end{aligned}
\end{equation}
where $\overline{P}_{SR,EH}$ and $\overline{P}_{DR,EH}$ denote the average harvested energy of S$\rightarrow$R and R$\rightarrow$D link, respectively, given by $\overline{P}_{SR,EH}=\frac{\eta \theta P_{S}\Omega _{sr,l}}{d_{sr}^{\alpha }}$, $\overline{P}_{DR,EH}=\frac{\eta \theta P_{D}\Omega _{rd,l}}{d_{rd}^{\alpha }}$, and $K_1$, $K_2$ are given by (\ref{eq:49func}) and (\ref{eq:50func}) respectively.

\textit{Proof}: Please refer to Appendix B.

\textbf{Case}~2: The S$\rightarrow$R link is selected when the buffer is empty, and in the next time slot, the decoded packet is transmitted from R to D when the buffer is full. The joint probability of this case is $P_{empty,P1}P_{full,P1}$. Therefore, one obtains $P_2$ as
\begin{equation}
{P_2} = {P_{empty,P1}}{P_{full,P1}}\left( {P_{e, SR}^{''} + P_{e, RD}^{''}} \right),
\label{eq:19func}
\end{equation}
where $P_{e, SR}^{''}$ and $P_{e, RD}^{''}$ are given by
\begin{equation}
\begin{aligned}
\hspace{-5mm}P_{e,SR}^{''} &\le E \left\{ {\frac{{\left( {1 - {P_{ES}}} \right)}}{2}erfc\left( {{{\left[ {\frac{8}{{{\gamma _{SR}}}} + \frac{{8\beta }}{{\gamma _{SR}^2}}} \right]}^{ - \frac{1}{2}}}} \right)} \right\}\\
&\hspace{-5mm}\approx \frac{{\left( {1 - {P_{ES}}} \right)}}{2}\sum\limits_{m = 1}^M {{\omega _m}} erfc\left( {\frac{{{e^{{\kappa _m}}}}}{{\sqrt {8{e^{{\kappa _m}}} + 8\beta } }}} \right)\frac{{\exp \left( {\kappa _{_m}^2} \right)}}{{\Gamma \left( {{L_{sr}}} \right)\overline \gamma  _{SR}^{{L_{sr}}}}}\\
 &\hspace{-5mm}\times \exp \left( {- \frac{{{e^{{\kappa _m}}}}}{{{{\overline \gamma  }_{SR}}}} + {\kappa _m}{L_{sr}}} \right) + {O_M},
 \label{eq:20func}
\end{aligned}
\end{equation}
\begin{equation}
\begin{aligned}
\hspace{5mm}P_{e,RD}^{''} &\le E \left\{ {\frac{1}{2}erfc\left( {{{\left[ {\frac{8}{{{\gamma _{RD}}}} + \frac{{8\beta }}{{\gamma _{RD}^2}}} \right]}^{ - \frac{1}{2}}}} \right)} \right\}\\
 &\hspace{-5mm}\approx  \frac{{N_{0,rd}^{{L_{sr}}}}}{{2{A^{{L_{sr}}}}\Gamma \left( {{L_{sr}}} \right)\Gamma \left( {{L_{rd}}} \right)\!\Omega _{rd,l }^{{L_{sr}}}\overline P _R^{{L_{sr}}}}}\sum\limits_{m = 1}^M {{\omega _m}erfc\left( {\frac{{{e^{{\kappa _m}}}}}{{\sqrt {8{e^{{\kappa _m}}} + 8\beta } }}} \right)} \\
& \hspace{-5mm}\times  \left( {{e^{{\kappa _m}{L_{sr}}}}\mathop G\nolimits_{0,2}^{2,0} \left( {\left. {\frac{{{N_{0,rd}}{e^{{\kappa _m}}}}}{{A{{\overline P }_R}{\Omega _{rd,l }}}}} \right|\begin{array}{*{20}{c}}
 - \\
{{L_{rd}} - {L_{sr}},0}
\end{array}} \right)\exp (\kappa_m^2)} \right) + {O_M}.
\label{eq:21func}
\end{aligned}
\end{equation}

\textit{Proof}: Please refer to Appendix B.

\textbf{Case}~3: In this case, the signal is transmitted from S to R when the buffer is empty. However, the buffer is not full when the R$\rightarrow$D link is selected. The joint probability of this case is $P_{empty,P1}(1-P_{full,P1})$. Hence, one obtains $P_3$ as
\begin{equation}
{P_3} = {P_{empty,P1}}\left( {1 - {P_{full,P1}}} \right)\left( {P_{e, SR}^{''} + P_{e, RD}^{'}} \right).
\label{eq:22func}
\end{equation}

\textbf{Case}~4: In this case, the buffer is not empty when the S$\rightarrow$R link is selected. However, when the decoded packet is transmitted to D, the buffer is full. The joint probability of this case is $(1-P_{empty,P1})P_{full,P1}$. Then, one obtains $P_4$ as
\begin{equation}
{P_4} = \left( {1 - {P_{empty,P1}}} \right){P_{full,P1}}\left( {P_{e, SR}^{'} + P_{e, RD}^{''}} \right).
\label{eq:23func}
\end{equation}

Here, combining the above four events, we can get the end-to-end BER, i.e., $P_{e1} =\sum_{f =1}^{4}P_{f}$. According to \cite{6613630}, the asymptotic upper bound on BER of the Protocol~$1$ can be obtained, as
\begin{equation}
\begin{aligned}
{P_{e1}} &\le  \left( {1 - {P_{empty,P1}}} \right)\left( {1 - {P_{full,P1}}} \right)\left( {P_{e, SR}^{'} + P_{e, RD}^{'}} \right)\\
 &+ {P_{empty,P1}}{P_{full,P1}}\left( {P_{e, SR}^{''} + P_{e, RD}^{''}} \right)\\
 &+ {P_{empty,P1}}\left( {1 - {P_{full,P1}}} \right)\left( {P_{e, SR}^{''} + P_{e, RD}^{'}} \right)\\
 &+ \left( {1 - {P_{empty,P1}}} \right){P_{full,P1}}\left( {P_{e, SR}^{'} + P_{e, RD}^{''}} \right),
 \label{eq:24func}
\end{aligned}
\end{equation}
where $P_{empty,P1}$ and $P_{full,P1}$ are given in (\ref{eq:41func}) and (\ref{eq:42func}), respectively.

\subsection{BER of the Protocol~$2$}
For the Protocol~$2$, the information transmission only depends on harvested energy and energy-shortage status, i.e., the packets can not be forced to transmit when the buffer is full (or empty). Therefore, the error probability can be divided into two cases: a) For $0\leq \phi  _{j}<J$, the S$\rightarrow$R link is selected only when $P_{SR,EH}>\delta P_{DR,EH}$ \& $P_{EH}>P_I$ and b) For $0< \phi  _{j}\leq J$, the R$\rightarrow$D link is selected only when $P_{SR,EH}<\delta P_{DR,EH}$. Based on the two events, the BER of the Protocol~$2$ is formulated as
\begin{equation}
\begin{aligned}
&{P_{e2}} = \frac{{{K_1}\overline P _{SR,EH}^{{L_{rd}}}}}{{\Gamma \left( {{L_{rd}}} \right)}}\sum\limits_{l = 0}^{{L_{sr}} - 1} {\frac{{{\delta ^l}\overline P _{DR,EH}^{l}\Gamma \left( {{L_{rd}} + l} \right)}}{{l!{{\left( {\delta {{\overline P }_{DR,EH}} + {{\overline P }_{SR,EH}}} \right)}^{l + {L_{rd}}}}}}}
 + \frac{{{K_2}{\delta ^{{L_{sr}}}}\overline P _{DR,EH}^{{L_{sr}}}}}{{\Gamma \left( {{L_{sr}}} \right)}}\sum\limits_{l = 0}^{{L_{rd}} - 1} {\frac{{\Gamma \left( {{L_{sr}} + l} \right)\overline P _{SR,EH}^{l}}}{{l!{{\left( {{{\overline P }_{SR,EH}} + \delta {{\overline P }_{DR,EH}}} \right)}^{l + {L_{sr}}}}}}}.
 \label{eq:25func}
\end{aligned}
\end{equation}

\begin{figure*}[tbp]
\center\vspace{-0.5cm}
\subfigure[State transition diagram of the Protocol~$1$]{
\includegraphics[width=3.4in,height=1.0in]{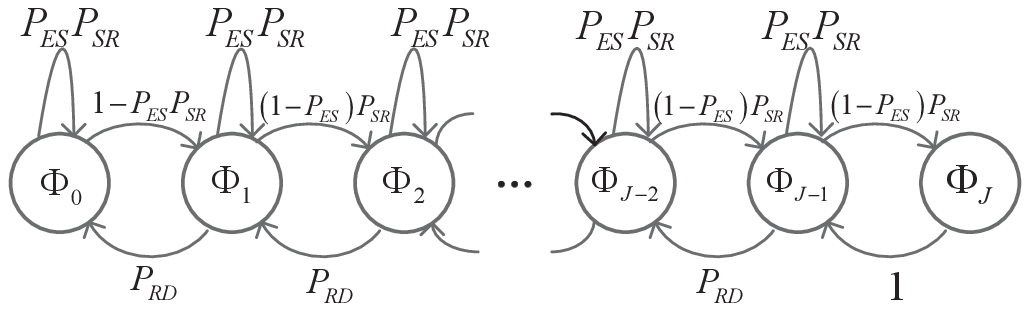}
\vspace{-0.5cm}}
\label{fig:subfig3a}
\subfigure[State transition diagram of the Protocol~$2$]{
\includegraphics[width=4.4in,height=1.0in]{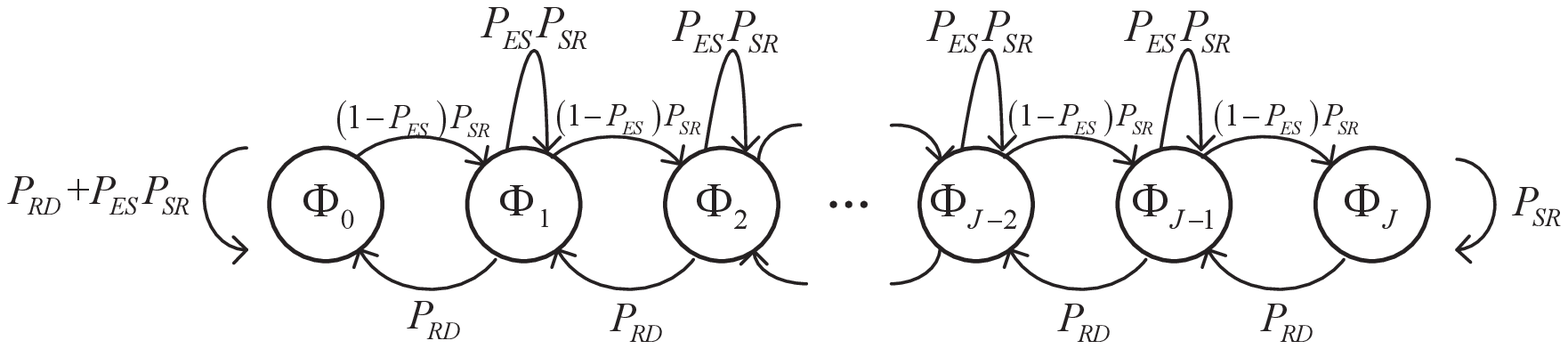}
\vspace{-0.5cm}}
\label{fig:subfig3b}
\centering
\caption{State diagram of the Markov chain for the status of the buffer at R.}
\label{fig:fig3}\vspace{-0.5cm}  
\end{figure*}

\section{Delay Analysis} \label{sect:delay analysis}
In this section, the average-delay closed-form expressions of the two protocols for the buffer-aided DCSK-SWIPT relay system are derived.
Fig.~\ref{fig:fig3} shows the state transition diagram at R for both protocols, where $P_{SR}$ and $P_{RD}$ denote the probabilities of selecting the S$\rightarrow$R and R$\rightarrow$D links, respectively.

\subsection{Protocol~$1$}
\subsubsection{Queuing Delay}
Based on the link-selection schemes, it can be concluded that the queuing delay is caused by the buffer at R. The delay of the packets and the queue length at the time slot $j$ are denoted as $T_{qt}^{'}\left ( j \right )$ and $Q_{qt}^{'}\left ( j \right )$, respectively. According to \cite{6613630}, the average delay $T_{qt}^{'}=E\left \{ T_{qt}^{'}\left ( j \right ) \right \}$, which denotes the average time that a packet is stored in the buffer, can be expressed as
\begin{equation}
T_{qt}^{'} = \frac{{Q_{qt}^{'}}}{{R_{qt}^{'}}},
\label{eq:26func}
\end{equation}
where $R_{qt}^{'}$ (in packets/slot) denotes the average arrival rate into the queue and $Q_{qt}^{'}=E \left \{ Q_{qt}^{'}\left ( j \right ) \right \}$ (in packets) denotes the average queue length at the buffer. Generally, only one packet is transmitted at each time slot, i.e., $R_{qt}^{'} = 1$ (packet/slot).


For the buffer size $J$, the average queue length can be written as
\begin{equation}
E \left\{ {Q_{qt}^{'}(j)} \right\} = \sum\limits_{j = 0}^J {j{P_{{\phi  _j}}}} ,
\label{eq:27func}
\end{equation}
where $P_{\phi  _{j}}$ denotes the probability of different buffer statuses of Protocol~$1$, i.e., $\phi  _{j},j\in \left \{ 0,1,...,J \right \}$. It should be noted that $P_{full,P1} = P_{\phi_J}$ and $P_{empty,P1} = P_{\phi_0}$.
From (\ref{eq:38func}), one can obtain
\begin{equation}
{P_{{\phi  _j}}} = \frac{{{P_{RD} ^{J - j - 1}}}}{{{{(1 - P_{ES})}^{J - j}}{P_{SR} ^{J - j}}}}{P_{{\phi  _J}}},{\rm{    1}} \le j \le J - 1 .
\label{eq:28func}
\end{equation}

Using (\ref{eq:27func}), (\ref{eq:28func}) and [8, Eq.~(34)], one has
\begin{equation}
Q_{qt}^{'} = {P_{{\phi  _J}}}\left( {\frac{{{{\left( {\frac{{{P_{RD}}}}{{\left( {1 - {P_{ES}}} \right){P_{SR}}}}} \right)}^{J - 1}} - {\rm{1}}}}{{{P_{RD}}{{\left(  {1- \frac{{\left( {1 - {P_{ES}}} \right){P_{SR}}}}{{{P_{RD}}}}} \right)}^2}}} - \frac{{J - 1}}{{{P_{RD}}\left( {1- \frac{{\left( {1 - {P_{ES}}} \right){P_{SR}}}}{{{P_{RD}}}}} \right)}} + J} \right).
\label{eq:29func}
\end{equation}

From Fig.~\ref{fig:fig3}(a), the average arrival rate into the buffer of R is derived as
\begin{equation}
\begin{aligned}
R_{qt}^{'} &= \left( {1 - {P_{ES}}{P_{SR}}} \right){P_{{\phi  _0}}} + \sum\limits_{j = 1}^{J - 1} {\left( {1 - {P_{ES}}} \right){P_{SR}}{P_{{\phi  _j}}}}\\
 &= (1 - {P_{ES}}){P_{SR}}\left( {1 - {P_{{\phi  _J}}}} \right) + {P_{RD}}{P_{{\phi  _0}}}.
\label{eq:30func}
\end{aligned}
\end{equation}

Based on (\ref{eq:26func}), (\ref{eq:29func}) and (\ref{eq:30func}), the closed-form expression of queuing delay is formulated as
\begin{equation}
\begin{aligned}
&T_{qt}^{'}= \frac{{{P_{{\phi  _J}}}}}{{\left( {1 - {P_{ES}}} \right){P_{SR}}\left( {1 - {P_{{\phi  _J}}}} \right) + {P_{RD}}{P_{{\phi  _0}}}}}
\hspace{-2mm}&\times \left( {\frac{{{{\left( {\frac{{{P_{RD}}}}{{\left( {1 - {P_{ES}}} \right){P_{SR}}}}} \right)}^{J - 1}} - {\rm{1}}}}{{{P_{RD}}{{\left( {1 - \frac{{\left( {1 - {P_{ES}}} \right){P_{SR}}}}{{{P_{RD}}}}} \right)}^2}}} - \frac{{J - 1}}{{{P_{RD}}\left( {1 - \frac{{\left( {1 - {P_{ES}}} \right){P_{SR}}}}{{{P_{RD}}}}} \right)}} + J} \right).
\label{eq:31func}
\end{aligned}
\end{equation}

\subsubsection{Silent time slots}
For the Protocol~$1$, the silent time slot is caused when the buffer is empty and the energy is insufficient to recover the received signal.
If $P_{SR,EH}< P_I$, the data buffer status is unchanged, but it also contributes to end-to-end delay.
To calculate the average number of time slots when the buffer remains empty before R receives a packet, we consider the sequence of the buffer status \{empty~\&~shortage, empty~\&~shortage,...,empty~\&~not shortage\}. Here, the number of time slots follows a geometric distribution with mean value ${{{P_{ES}}}}/{{(1 - {P_{ES}})}}$. Hence, the average delay of the silent time slots can be computed as
\begin{equation}
T_{st}^{'} = \frac{{\gamma \left( {{L_{sr}},\frac{{{L_{sr}}{P_I}d_{sr}^\alpha }}{{\eta \theta {P_S}}}} \right)}}{{\Gamma \left( {{L_{sr}}} \right) - \gamma \left( {{L_{sr}},\frac{{{L_{sr}}{P_I}d_{sr}^\alpha }}{{\eta \theta {P_S}}}} \right)}}.
\label{eq:32func}
\end{equation}

Finally, combining (\ref{eq:31func}) and (\ref{eq:32func}), the closed-form expression of average delay can be obtained as
\begin{equation}
\begin{aligned}
&  T_{t1}^{'} = T_{qt}^{'} + T_{st}^{'}\\
& \hspace{-1mm} = \left( {\frac{{{{\left( {\frac{{{P_{RD}}}}{{\left( {1 - {P_{ES}}} \right){P_{SR}}}}} \right)}^{J - 1}} - {\rm{1}}}}{{{P_{RD}}{{\left( {1 - \frac{{\left( {1 - {P_{ES}}} \right){P_{SR}}}}{{{P_{RD}}}}} \right)}^2}}} - \frac{{J - 1}}{{{P_{RD}}\left( {1 - \frac{{\left( {1 - {P_{ES}}} \right){P_{SR}}}}{{{P_{RD}}}}} \right)}} + J} \right)\\
 &\hspace{-1mm}  \times\frac{{{P_{{\phi  _J}}}}}{{\left( {1 - {P_{ES}}} \right){P_{SR}}\left( {1 - {P_{{\phi  _J}}}} \right) + {P_{RD}}{P_{{\phi  _0}}}}} + \frac{{\gamma \left( {{L_{sr}},\frac{{{L_{sr}}{P_I}d_{sr}^\alpha }}{{\eta \theta {P_S}}}} \right)}}{{\Gamma \left( {{L_{sr}}} \right) - \gamma \left( {{L_{sr}},\frac{{{L_{sr}}{P_I}d_{sr}^\alpha }}{{\eta \theta {P_S}}}} \right)}}.
\end{aligned}
 \label{eq:33func}
\end{equation}

\subsection{Protocol~$2$}
For the Protocol~$2$, we do not force S (or R) to transmit if the buffer is full (or empty), which in turn results in silent time slots. In this time slot, although no transmission takes place, it also contributes to the end-to-end packet delay. To compute the average delay, both the queuing delay and the delay of the silent time slots are taken into account. Similarly, the probabilities of the buffer being full and empty for the Protocol~$2$, which are defined as $P_{full,P2}$ and $P_{empty,P2}$, respectively. Also, the closed-form expressions of $P_{full,P2}$ and $P_{empty,P2}$ are derived in Appendix A.
\subsubsection{Queuing delay}
In the following, the queuing delay is derived based on the proposed link-selection protocol.
The probability of different buffer statuses of the Protocol~$2$ is defined as $P_{\Psi  _{j}}$, i.e., $\Psi   _{j},j\in \left \{ 0,1,...,J \right \}$, where $P_{full,P2} = P_{\Psi _J}$ and $P_{empty,P2} = P_{\Psi _0}$.
According to the Eq.~\eqref{eq:27func}, the average queue length $Q_{qt}^{''}=E \left \{ Q_{qt}^{''}(j) \right \}$ is given by
\begin{equation}
Q_{qt}^{''} = {P_{{ \Psi   _J}}}\left[ {\frac{{{{\left( {\frac{{{P_{RD}}}}{{(1 - {P_{ES}})(1 - {P_{RD}})}}} \right)}^{J - 1}} - 1}}{{{{\left( {1 \- \frac{{(1 - {P_{ES}})(1 - {P_{RD}})}}{{{P_{RD}}}}} \right)}^2}}} - \frac{{J - 1}}{{1 - \frac{{(1 - {P_{ES}})(1 - {P_{RD}})}}{{{P_{RD}}}}}} + J} \right].
\label{eq:34func}
\end{equation}

According to Fig.~\ref{fig:fig3}(b), the average arrival rates can be expressed as
\begin{equation}
\begin{aligned}
R_{qt}^{''} = \sum\limits_{j = 0}^{J - 1} {(1 - {P_{ES}}){P_{SR}}{P_{{\Psi  _j}}}}  = (1 - {P_{ES}}){P_{SR}}\left( {1 - {P_{{\Psi  _J}}}} \right).
 \label{eq:35func}
\end{aligned}
\end{equation}

Using (\ref{eq:34func}) and (\ref{eq:35func}), the queuing delay of the Protocol~$2$ can be calculated by $T_{qt}^{''}=Q_{qt}^{''}/R_{qt}^{''}$.

\subsubsection{Silent time slots}
According to Table II, there are two cases when a silent time slot occurs: Case A) when the buffer is empty, the R$\rightarrow$D link is selected, and Case B) when the buffer is full and $P_{SR,EH}>\delta P_{DR,EH}$, but the harvested energy is insufficient to recover the received signal, i.e., $P_{SR,EH}<P_I$. For Case A, the buffer status, i.e., \{empty, empty, ..., not empty\}, is considered to calculate the average number of time slots that the buffer remains empty before R receives a packet. For Case B, the average delay of the silent time slots is the same as the Protocol~$1$. In the following, the delay caused by Case A is derived as
\begin{equation}
\begin{aligned}
&T_{cs}^{''} = \frac{{{P_{{\Psi  _0}}}}}{{1 - {P_{{\Psi  _0}}}}}\\
 &= \frac{{{P_{RD}^J}\left( {{P_{RD}} - (1 - {P_{ES}})(1 - {P_{RD}})} \right)}}{{(1 - {P_{ES}})(1 - {P_{RD}})\left( {{P_{RD}^J} - {{\left( {(1 - {P_{ES}})(1 - {P_{RD}})} \right)}^J}} \right)}},
 \label{eq:36func}
\end{aligned}
\end{equation}
where $P_{\Psi  _{0}}$ is given by (\ref{eq:44func}).

Combining (\ref{eq:32func}), (\ref{eq:34func}), (\ref{eq:35func}) and (\ref{eq:36func}), the closed-form expression of the average delay can be obtained as
\begin{equation}
\begin{aligned}
&T_{t2}^{''} = T_{qt}^{''} + T_{cs}^{''} + T_{st}^{'}\\
 & \hspace{-1mm} = \frac{{{P_{{\Psi  _J}}}}}{{(1 - {P_{ES}}){P_{SR}}\left( {1 - {P_{{\phi  _J}}}} \right)}}\left[ {\frac{{{{\left( {\frac{{{P_{RD}}}}{{(1 - {P_{ES}})(1 - {P_{RD}})}}} \right)}^{J - 1}} - 1}}{{{{\left( {1 - \frac{{(1 - {P_{ES}})(1 - {P_{RD}})}}{{{P_{RD}}}}} \right)}^2}}}} \right.\\
 &\hspace{-1mm}- \left. {\frac{{J - 1}}{{1 - \frac{{(1 - {P_{ES}})(1 - {P_{RD}})}}{{{P_{RD}}}}}} + J} \right] + \frac{{\gamma \left( {{L_{sr}},\frac{{{L_{sr}}{P_I}d_{sr}^\alpha }}{{\eta \theta {P_S}}}} \right)}}{{\Gamma \left( {{L_{sr}}} \right) - \gamma \left( {{L_{sr}},\frac{{{L_{sr}}{P_I}d_{sr}^\alpha }}{{\eta \theta {P_S}}}} \right)}}\\
 &\hspace{-1mm}+ \frac{{{P_{RD}}^J\left( {{P_{RD}} - (1 - {P_{ES}})(1 - {P_{RD}})} \right)}}{{(1 - {P_{ES}})(1 - {P_{RD}})\left( {{P_{RD}}^J - {{\left( {(1 - {P_{ES}})(1 - {P_{RD}})} \right)}^J}} \right)}}.
 \label{eq:37func}
\end{aligned}
\end{equation}

\section{Simulation Results And Discussions} \label{sect:SIMU}
In this section, numerical results in terms of BER and average delay are discussed to demonstrate the superiority of the proposed link-selection protocols for the buffer-aided DCSK-SWIPT relay system. In simulations, both the transmit powers of S and D are equal, i.e., $P_S=P_D=1$~dBm. The spreading factor is $\beta=160$ and the energy conversion efficiency factor $\eta$ is set to 0.6. To match the realistic scenario, we assume that the energy of decoding cost is $P_I=\frac{P_S}{100}$. Moreover, the antenna and circuit noise are assumed to have equal variances, i.e., $N_{0,sr}=N_{0,rd}=N_{0,IR}=N_0$. For multipath Rayleigh channels, three-path channels are considered, and the channel parameters are set to $E\left \{ h _{1}^{2} \right \}=E\left \{ h _{2}^{2} \right \}=E\left \{ h _{3}^{2} \right \}=\frac{1}{3}$, $\tau _{1}=0,\tau _{2}=2,\tau _{3}=5$ for all links. Unless otherwise noted, $\theta$ and $\delta$ are set to $0.5$ and $1.05$, respectively.

\subsection{BER Performance}
\begin{figure}[!tbp]
\center
\vspace{-0.2cm}
\includegraphics[width=3.2in,height=2.2in]{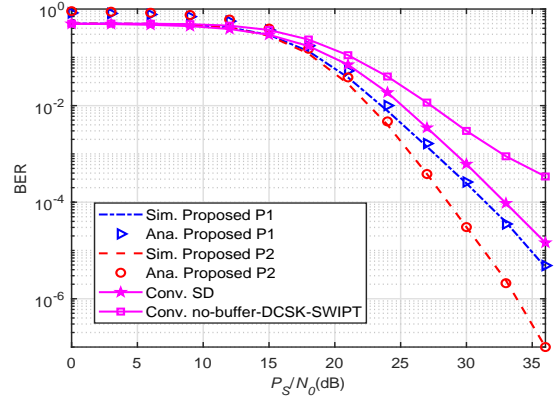}
\vspace{-0.2cm}
\caption{The BER performance curves of different systems over multipath Rayleigh fading channels, where the theoretical and simulated results of the proposed protocols are verified.}
\label{fig:fig4}
\end{figure}

Fig.~\ref{fig:fig4} shows the BER performance for the proposed system (two link-selection protocols), the DCSK-SWIPT cooperative system (Conv-no-buffer-DCSK-SWIPT) and conventional DCSK system (Conv-SD) over multipath Rayleigh fading channels, where the buffer size is set to $J=10$ .
It can be observed that the simulated BER curves well match with the theoretical ones, which verifies the correctness of the proposed analytical method.
Moreover, the proposed system can obtain better performance compared with the DCSK-SWIPT cooperative system and the conventional DCSK system.
In addition, for the proposed system, the Protocol~$2$ can achieve superior BER performance compared with the Protocol~$1$. For example, at a BER of $10^{-5}$, the Protocol~$1$ can achieve a $2$ dB gain compared with the DCSK-SWIPT cooperative system while the Protocol~$2$ can obtain a $3$~dB gain in comparison to the Protocol~$1$.

\begin{figure}[!tbp]
\center
\vspace{-0.35cm}
\includegraphics[width=3.2in,height=2.2in]{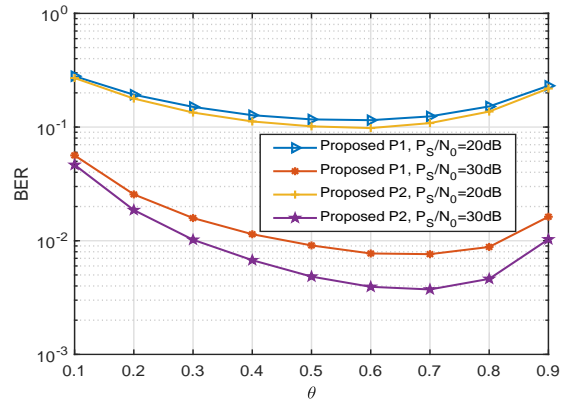}
\vspace{-0.2cm}
\caption{BER versus power splitting ratio $\theta$ for the two proposed protocols at $\frac{P_{S}}{N_{0}} = 20$~dB and $\frac{P_{S}}{N_{0}} = 30$~dB, where $d_{sr}=d_{rd}=1$.}  
\label{fig:fig5}
\end{figure}

In Fig.~\ref{fig:fig5}, the BER curves are plotted against the power splitting ratio $\theta$ for the proposed system with different SNR.
It is observed that better BER performance can be obtained when $\theta$ increases from $0.1$ to $0.6$ while
it gets worse rapidly if $\theta>0.7$.
The reason is that as $\theta$ increases from $0.1$ to $0.6$, more power is used for harvested energy, thus the energy shortage probability becomes lower.
However, when the harvested energy is sufficient to recover the modulated signal, the energy used for information transmission decreases when the value of $\theta$ increases continuously. It means that part of the power used for information decoding is redundant when $\theta$ exceeds the optimal values.
Hence, the BER performance of both protocols reach an optimal value at $\theta=0.65$ because the proposed system can achieve better tradeoff between information transmission and harvested energy.

\begin{figure}[!tbp]
\vspace{-0.45cm}
\center
\vspace{-0.5cm}
\includegraphics[width=3.2in,height=2.2in]{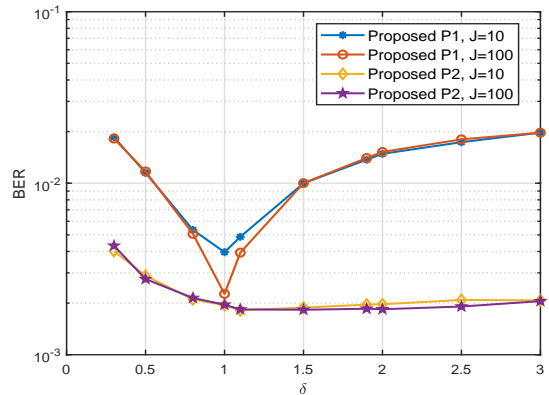}
\vspace{-0.5cm}
\caption{BER versus $\delta$ for different buffer size, where $\frac{P_{S}}{N_{0}}=25$~dB and $d_{sr}=d_{rd}=1$.}  
\label{fig:fig6}
\end{figure}
\begin{figure}[!tbp]
\center\vspace{-0.5cm}
\includegraphics[width=3.2in,height=2.2in]{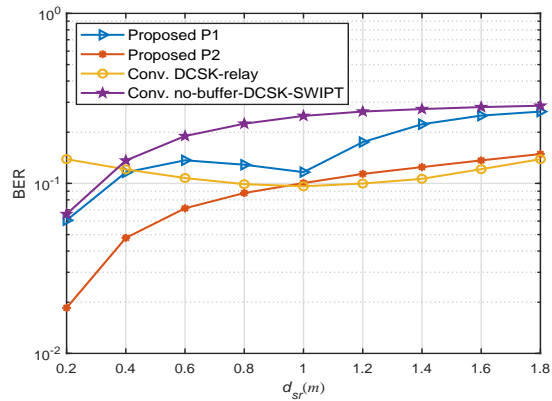}
\vspace{-0.5cm}
\caption{BER versus $d_{sr}$ for different systems at $\frac{P_{S}}{N_{0}} = 20$~dB, where $L_{sr}=L_{rd}=1$, $d_{sr} + d_{rd} = 2$.}
\label{fig:fig7}
\vspace{-0.5cm}
\end{figure}

Fig.~\ref{fig:fig6} plots the BER versus the threshold $\delta$ for different buffer size. For Protocol~$1$, it can be observed that $\delta=1$ results in the lowest BER in all cases, which is similar to the conventional DCSK system. However, for Protocol~$2$, the BER decreases with the $\delta$ increases and the BER tends to be steady when $\delta>2$.
This is because the Protocol~$2$ does not force S and R to transmit information in case of poor link quality, which is different from the Protocol~$1$.

\begin{figure}[htb]
\centering
\vspace{-0.3cm}
\subfigure[\hspace{-0.8cm}]{ \label{fig:subfig:8a} 
\includegraphics[width=3.2in,height=2.2in]{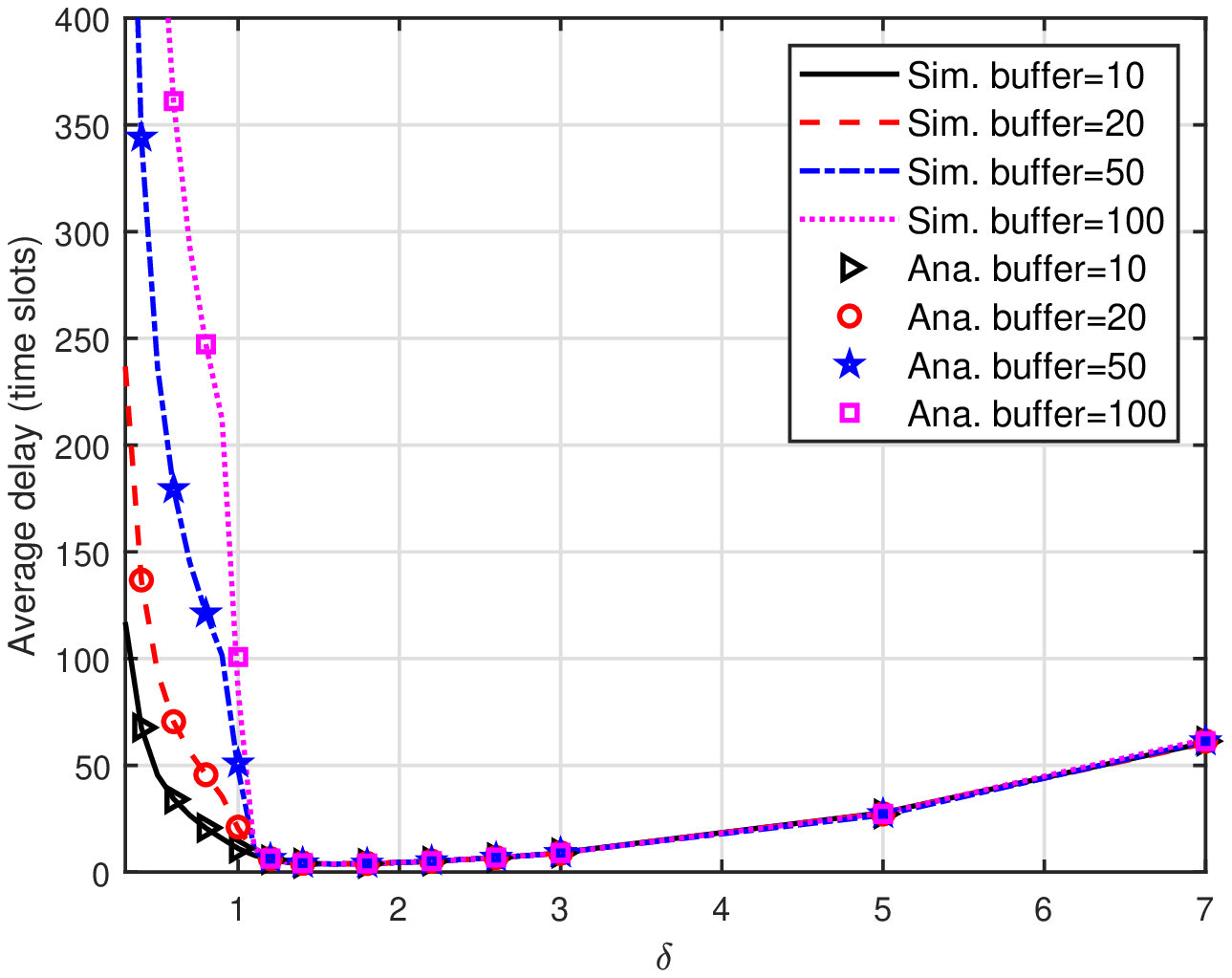}}
\subfigure[\hspace{-0.8cm}]{ \label{fig:subfig:8b} 
\includegraphics[width=3.2in,height=2.2in]{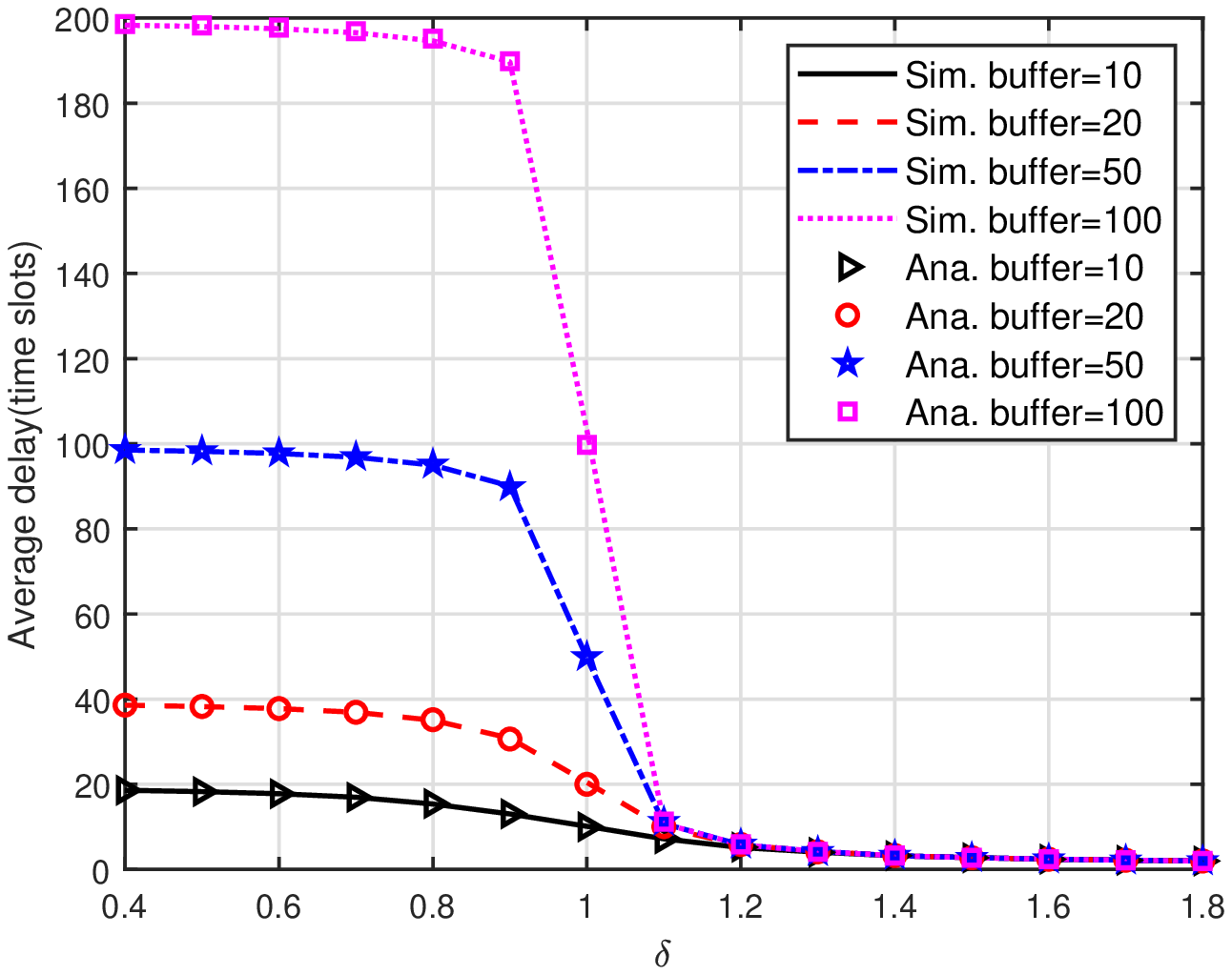}}
\caption{Average delay vs. $\delta $ of (a)~Protocol~$1$ and (b)~Protocol~$2$, where $\frac{P_{S}}{N_{0}} = 30$~dB and $J=10$.}
\label{fig:fig8} 
\vspace{-0.45cm}
\end{figure}

In Fig.~\ref{fig:fig7}, the BER curves of the proposed system, DCSK-SWIPT cooperative system (Conv-no-buffer-DCSK-SWIPT) and conventional DCSK relay system (Conv-DCSK-relay) are plotted against the distance between S and R, i.e., $d_{sr}$.
It shows the similar results that the BER performance gets worse with the increasing of $d_{sr}$. This is because, when R moves away from S, the probability of successful decoding and the value of harvested energy at R decreases due to the influence of increasing path loss. Moreover, as the R moves closer to D, the decreasing of R$\rightarrow$D distance causes lower path loss, thus the reliable communication between R and D can be guaranteed only by consuming lower harvested energy.
It explains the phenomenon that the BER performance tends to be steady when $d_{sr}>1.4m$.
Moreover, the BER of the Protocol~$1$ decreases rapidly when $d_{sr}=1$.
This phenomenon is caused by the value of $\theta$.
In addition, it can be observed that the conventional DCSK-SWIPT cooperative communication systems also have the same performance as compared with the Protocol~$2$, which is different from conventional DCSK relay system.

\subsection{Delay Performance}
Fig.~\ref{fig:fig8} shows the average delay versus the threshold $\delta $ for both protocols when different buffer sizes are considered.
Referring to this two figures, the average-delay curves from the theoretical analysis agree with the ones from the simulation, which shows the validation for the proposed analytical method.
Moreover, the following insights can be obtained:
1)~Protocol~$1$: The average delay decreases as the parameter $\delta $ increases from $0.2$ to $1.8$. This is because that with the increases of $\delta$ the R$\rightarrow$D link is more frequently selected. However, when $\delta$ surpasses the optimal value, lower average delay comes at the expense of BER performance because the link with better performance is not always chosen. Moreover, when $\delta>1.2$, the average delay decreases slowly. Furthermore, the average delay increases as the buffer size increases from $10$ to $100$.
2)~Protocol~$2$: As $\delta \rightarrow0$, the average delay increases sharply since the S$\rightarrow$R link is more frequently chosen. This case results in higher average queuing delay. However, the value of average delay for different buffer size increases slowly when $\delta >2$, this is because that the static time slots increase as R$\rightarrow$D link is more frequently selected. Similarly, the average delay of the Protocol~$2$ also increases with the increases of the buffer size.

\begin{figure}[!tbp]
\center
\vspace{-0.4cm}
\includegraphics[width=3.2in,height=2.2in]{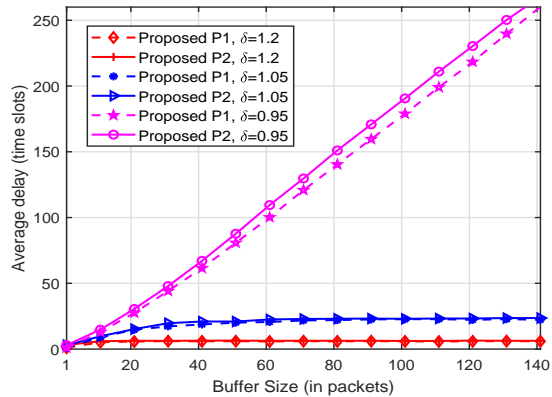}
\vspace{-0.2cm}
\caption{Average delay vs. buffer size for the Protocol~$1$ and Protocol~$2$.}  
\label{fig:fig9}
\end{figure}

Fig.~\ref{fig:fig9} shows the average delay versus buffer size $J$ for different threshold $\delta $. For both the proposed protocols, the average delay increases rapidly as the buffer size increases. This is because the energy shortage probability decreases with increasing harvested energy. Furthermore, the average delay of the Protocol~$1$ is almost the same as that of Protocol~$2$ for three different values of $\delta$.

\begin{figure}[!tbp]
\center
\vspace{-0.3cm}
\includegraphics[width=3.2in,height=2.2in]{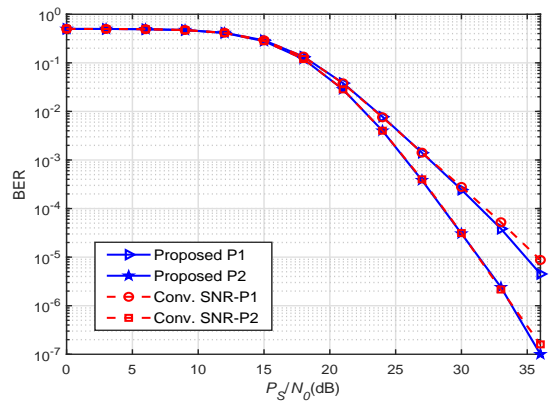}
\vspace{-0.2cm}
\caption{BER of the proposed buffer-aided DCSK SWIPT relay system with two protocols based on harvested energy and SNR.}  
\label{fig:fig10}
\end{figure}

\subsection{Performance Comparison Between the Proposed Strategies and the Conventional SNR-based ones}
\begin{figure}[!tbp]
\center
\vspace{-0.3cm}
\includegraphics[width=3.2in,height=2.2in]{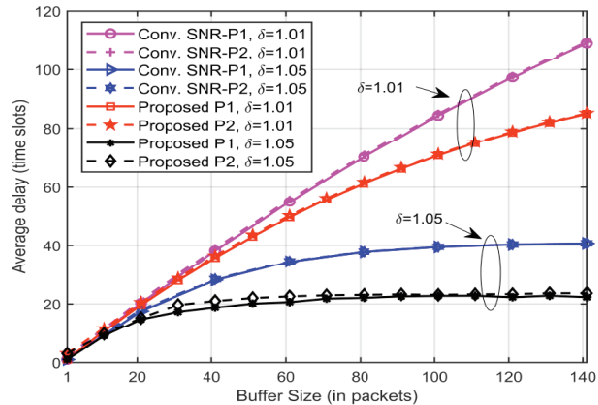}
\vspace{-0.2cm}
\caption{Average delay of the proposed buffer-aided DCSK-SWIPT relay system with two protocols based on harvested energy and SNR.}  
\label{fig:fig11}
\end{figure}

To further show the effectiveness of the proposed protocols, the performance comparison between the proposed strategies and the conventional SNR-based ones is given in this section.
Figs.~\ref{fig:fig10} and \ref{fig:fig11} show BER and average delay of the proposed buffer-aided DCSK-SWIPT relay system with two protocols based on harvested energy and SNR, respectively.
As can be seen from Fig.~\ref{fig:fig10}, the BER performance of the proposed protocols based on harvested energy can be improved compared with the conventional SNR-based ones.
Also, it can be observed from Fig.~\ref{fig:fig11} that the average delay of the proposed protocols are lower than that of the conventional SNR-based ones.
The reason is that the conventional SNR-based protocols need to perform the channel estimation and result in the energy shortage, thus decreasing BER performance and increasing the delay.
Moreover, for the conventional SNR-based strategies, the channel estimator not only leads to high complexity of the nodes but also consumes more energy at the relay.
Hence, it implies that the proposed protocols can be considered as an outstanding alternative to the SNR-based ones due to better BER performance, lower average delay and more simple implementation.

\section{Conclusion} \label{sect:conclusion}
In this paper, a buffer-aided DCSK-SWIPT relay system has been investigated over multipath Rayleigh fading channels.
Moreover, two link-selection protocols have been formulated, which can be easily realized without acquiring CSI. The practical problem of the decoding cost at the relay has also been considered.
Furthermore, the closed-form expressions of the BER and average-delay for the proposed system have been derived over multipath Rayleigh fading channels.
Both theoretical analysis and simulations have demonstrated that the proposed protocols not only achieve better performance in terms of BER and average delay but also have lower hardware complexity compared with the existing SNR-based ones. Also, results show that both protocols can offer better BER performance than the conventional DCSK system and DCSK-SWIPT relay system.
Thanks to the aforementioned advantages, the proposed buffer-aided DCSK-SWIPT relay system can be considered as a excellent alternative for low-power and low-complexity short-range wireless communication environments.

\begin{appendices}
\section{Probability Of Full And Empty Buffer}
In this appendix, the probabilities that the buffer is empty or full for the Protocol~$1$ and Protocol~$2$ are derived. Here, we take the derivation of the Protocol~$1$ as an example and the similar method can be used in the Protocol~$2$. As shown in Table I, the data buffer status is determined by the harvested energy and the energy-shortage status at R. It should be noted that $P_{SR}+P_{RD}=1$ holds.

An Markov chain is used to represent the status of the buffer and the transmission process of packets between different time slots \cite{Papoulis1984Probability}. Let $\Phi _{\nu}=\nu,\nu=0,1,...,J$, denotes the $\nu$th state in the Markov chain, where $\nu$ denotes the number of information stored in the buffer. Actually, the buffer size $J$ has a Markov chain of $\Phi_{\nu}=J+1$ states. According to the link-selection criterion in Table I, one has the following three cases:

Case 1: If the buffer is empty, i.e., the buffer is in $\Phi_0$, there are two possible transitions. The first one is that the packet is successfully transmitted to R, and the status of the buffer is from $\Phi_0$ to $\Phi_1$ with the probability $p_{0,1}=1-P_{ES}P_{SR}$. The other one is that the energy is insufficient to recover the received signal, i.e., the outage of S$\rightarrow$R link occurs. At this time slot, the status of the buffer is just from $\Phi_0$ to $\Phi_0$, and the probability of this situation is $p_{0,0}=P_{ES}P_{SR}$.

Case 2: If the buffer is neither full nor empty, i.e, $\Phi _{\nu},\nu\in \left \{ 1,...,J-1 \right \}$, the S$\rightarrow$R and R$\rightarrow$D links are selected based on the harvested energy and data-buffer status. Similarly, there are two cases taking into account: a) If $P_{SR,EH}> \delta  P_{DR,EH} ~\&~ P_{SR,EH}>P_I$ holds, the signal is transmitted from S to R, and then the transition is from $\Phi_{\nu}$ to $\Phi_{\nu+1}$ with probability $p_{\nu,\nu+1}=(1-P_{ES})P_{SR}$. However, if $P_{SR,EH}>P_I$, the outage of S$\rightarrow$R link occurs, i.e., the transition is from $\Phi_{\nu}$ to $\Phi_{\nu}$ with probability $p_{\nu,\nu}=P_{ES}P_{SR}$, and b) If $P_{SR,EH}< \delta  P_{DR,EH}$ holds, the R$\rightarrow$D link is selected, thus the transition is from $\Phi_{\nu}$ to $\Phi_{\nu-1}$ with probability $p_{\nu,\nu-1}=P_{RD}$.

Case 3: If the buffer is full, i.e, the buffer is in $\Phi_J$. In this case, the R$\rightarrow$D link is selected for transmission without considering other conditions, thus the status of the buffer is from $\Phi_{\nu}$ to $\Phi_{\nu-1}$ with probability $p_{\nu,\nu-1}=P_{RD}$.

As depicted in Fig.~\ref{fig:fig3}(a), one obtains
\begin{equation}
\begin{array}{l}
{P_{{\Phi _0}}} = {P_{ES}}{P_{SR}}{P_{{\Phi _0}}} + {P_{RD}}{P_{{\Phi _1}}},\\
{P_{{\Phi _1}}} = \left( {1 - {P_{ES}}{P_{SR}}} \right){P_{{\Phi _0}}} + {P_{ES}}{P_{SR}}{P_{{\Phi _1}}} + {P_{RD}}{P_{{\Phi _2}}},\\
{P_{{\Phi _{ \nu }}}} = \left( {1 - {P_{ES}}} \right){P_{SR}}{P_{{\Phi _{ \nu  - 1}}}} + {P_{ES}}{P_{SR}}{P_{{\Phi _{ \nu }}}} + {P_{RD}}{P_{{\Phi _{ \nu  + 1}}}}{\rm{   }},\\
{P_{{\Phi _{J - 1}}}} = \left( {1 - {P_{ES}}} \right){P_{SR}}{P_{{\Phi _{J - 2}}}} + {P_{ES}}{P_{SR}}{P_{{\Phi _{J - 1}}}} + {P_{{\Phi _J}}},\\
{P_{{\Phi _J}}} = \left( {1 - {P_{ES}}} \right){P_{SR}}{P_{{\Phi _{J - 1}}}},
\label{eq:38func}
\end{array}
\end{equation}
where $2\le \nu \le J-2$ and $P_{\Phi_{\nu}}$ represents the probability in $\Phi_{\nu}$.

According to (\ref{eq:38func}), one has
\begin{equation}
\begin{aligned}
&{P_{{\Phi _{ \nu }}}} = \frac{{\left( {1 - {P_{ES}}{P_{SR}}} \right)P_{SR}^{ \nu  - 1}{{\left( {1 - {P_{ES}}} \right)}^{ \nu  - 1}}}}{{P_{RD}^{\nu}}}{P_{{\Phi _0}}},{\rm{    }}1 \le  \nu  \le J - 1.
\label{eq:39func}
\end{aligned}
\end{equation}

Based on (\ref{eq:39func}), the relationship between $P_{\Phi_{0}}$ and $P_{\Phi_{J}}$ can be given by
\begin{equation}
{P_{{\Phi _0}}} = \frac{{P_{RD}^{J - 1}}}{{\left( {1 - {P_{ES}}{P_{SR}}} \right)P_{SR}^{J - 1}{{(1 - {P_{ES}})}^{J - 1}}}}{P_{{\Phi _J}}}.
\label{eq:40func}
\end{equation}

Combining (\ref{eq:39func}) with (\ref{eq:40func}) and using the fact that $\sum_{\nu=0}^{J}P_{\phi_{\nu}}=1$, one obtains the probability of the buffer being full and empty
\begin{equation}
{P_{full,P1}} = {\left[ {1 + \frac{1}{{{P_{RD}}}}\frac{{\xi  - {\xi ^J}}}{{1 - \xi }} + \frac{{{\xi ^{J - 1}}}}{{1 - P_{ES}{P_{SR}}}}} \right]^{ - 1}},
\label{eq:41func}
\end{equation}
and
\begin{equation}
{P_{empty,P1}} = \frac{{{\xi ^{J - 1}}}}{{1 - P_{ES}{P_{SR}}}}{P_{full,P1}},
\label{eq:42func}
\end{equation}
where $\xi =\frac{P_{RD}}{\left ( 1-P_{ES} \right )P_{SR}}$, and the probability for selecting S$\rightarrow$R and R$\rightarrow$D links are given by $P_{SR}=\Pr\left ( P _{SR,EH}> \delta  P_{DR,EH} \right )$ and $P_{RD}=\Pr\left ( P_{DR,EH}> P_{SR,EH}/\delta \right )$ respectively, derived in Appendix C.

Similarly, according to the Table II and Fig.~\ref{fig:fig3}(b), the probabilities of the buffer being full and empty of Protocol~$2$ can be computed as
\begin{equation}
\hspace{5mm}{P_{full,P2}} = \frac{{P_{SR}^J{{\left( {1 - {P_{ES}}} \right)}^J}\left( {{P_{SR}}\left( {1 - {P_{ES}}} \right) - {P_{RD}}} \right)}}{{P_{SR}^{J + 1}{{\left( {1 - {P_{ES}}} \right)}^{J + 1}} - P_{RD}^{J + 1}}},
\label{eq:43func}
\end{equation}
\begin{equation}
\hspace{-1mm}{P_{empty,P2}} = \frac{{P_{RD}^J\left( {{P_{RD}} - (1 - {P_{ES}})(1 - {P_{RD}})} \right)}}{{P_{RD}^{J + 1} - {{\left( {(1 - {P_{ES}})(1 - {P_{RD}})} \right)}^{J + 1}}}}.
\label{eq:44func}
\end{equation}

\section{Some Approximated BER Expressions}
In this appendix, we derive BER expressions for $P_{_{e,SR}}^{'}$, $P_{_{e,RD}}^{'}$, $P_{e,SR}^{''}$ and $P_{e,RD}^{''}$ given in (\ref{eq:17func}), (\ref{eq:18func}), (\ref{eq:20func}) and (\ref{eq:21func}), respectively.
For the harvested energy of S and D, the PDF of $P_{SR,EH}$ and $P_{DR,EH}$ can be given by
\begin{equation}
\begin{aligned}
{f_{{P_{\varsigma ,EH}}}}(x) = \frac{{{x^{{L_\varsigma } - 1}}}}{{\Gamma \left( {{L_\varsigma }} \right)\overline P _{\varsigma ,EH}^{{L_\varsigma }}}}\exp \left( { - \frac{x}{{{{\overline P }_{\varsigma ,EH}}}}} \right),
\label{eq:45func}
\end{aligned}
\end{equation}
where $\varsigma \in \left \{ SR,RD \right \}$ and $\overline{P}_{\varsigma ,EH}^{L_{\varsigma }}$ denotes the average harvested energy of S (or D).

Conditioned on the event ${P_{SR,EH}}> \delta {P_{RD,EH}}$, the BER of the S$\rightarrow$R link is formulated as
\begin{equation}
\begin{aligned}
\hspace{-1mm}P_{_{e,SR}}^{'}& = E\left\{ {\left[ {\frac{{\left( {1 - {P_{ES}}} \right)}}{2}erfc\left( {{{\left[ {\frac{8}{{{\gamma _{SR}}}} + \frac{{8\beta }}{{\gamma _{SR}^2}}}\right]}^{ - \frac{1}{2}}}} \right)} \right]|{P_{SR,EH}} >\delta {P_{DR,EH}}} \right\}\\
\hspace{-2mm}& = \frac{{E\left\{ {\left[ {\frac{{\left( {1 - {P_{ES}}} \right)}}{2}erfc\left( {{{\left[ {\frac{8}{{{\gamma _{SR}}}} + \frac{{8\beta }}{{\gamma _{SR}^2}}} \right]}^{ - \frac{1}{2}}}} \right)} \right] \cap {P_{SR,EH}} > \delta {P_{DR,EH}}} \right\}}}{{\Pr ({P_{SR,EH}} > \delta {P_{DR,EH}})}}.
\label{eq:46func}
\end{aligned}
\end{equation}

One obtains the numerator of the right hand side of (\ref{eq:46func}) as
\begin{equation}
\begin{aligned}
\hspace{-1mm}{K_1} &= E \left\{ {\left[ {\frac{{\left( {1 - {P_{ES}}} \right)}}{2}erfc\left( {{{\left[ {\frac{8}{{{\gamma _{SR}}}} + \frac{{8\beta }}{{\gamma _{SR}^2}}} \right]}^{ - \frac{1}{2}}}} \right)} \right] \cap {P_{SR,EH}} > \delta {P_{DR,EH}}} \right\}\\
\hspace{-1mm}& = \int_0^\infty  {\int_0^{\frac{v}{\delta }} {\frac{{\left( {1 - {P_{ES}}} \right)}}{2}erfc\left( {{{\left[ {\frac{8}{{{W_1}v}} + \frac{{8\beta }}{{{{\left( {{W_1}v} \right)}^2}}}} \right]}^{ - \frac{1}{2}}}} \right)} }
 {f_{{P_{SR,EH}}}}(v){f_{{P_{RD,EH}}}}(\zeta)d\zeta dv,
\label{eq:47func}
\end{aligned}
\end{equation}
where $W_1=\frac{2\left ( 1-\theta  \right )}{\eta \theta N_{0,IR}}$.

Using [42, eq. (3.351)], Eq.~(\ref{eq:47func}) can be simplified to an one-integral formula.
Then, we evaluate the integral using the Gauss-Hermite quadrature approach \cite{Abramowitz}. After some mathematical manipulations, Eq.~(\ref{eq:47func}) can have a similar form and can be solved with the same methodology as follows
\begin{equation}
\begin{aligned}
\int_{ - \infty }^\infty  {g(u)du = \sum\limits_{m = 1}^M {{w_m}} }g({u_m}) {exp({u_m^2})} + O_M,
\label{eq:48func}
\end{aligned}
\end{equation}
where $M$ denotes the number of sample points used for approximation, $u_m$ denotes the $m^{th}$ root of the Hermite polynomial $H_M(u)(m = 1,2,...M)$, $w_m$ denotes the $m$-th associated weight given by $\frac{2^{M-1}M!\sqrt{\pi }}{M^{2}H_{M-1}^{2}\left ( u_{m} \right )}$ and $O_M$ is the remainder term which decreases to zero as $M$ tends to infinite. By subsisting (\ref{eq:45func}) into (\ref{eq:47func}), $\kappa=\ln(u)$ is used to get the new limits of the integral from $-\infty $ to $\infty $. Hence, Eq.~(\ref{eq:47func}) can be simplified as
\begin{equation}
\begin{aligned}
\hspace{-1mm}&{K_1} = E \left\{ {\left[ {\frac{{\left( {1 - {P_{ES}}} \right)}}{2}erfc\left( {{{\left[ {\frac{8}{{{\gamma _{SR}}}} + \frac{{8\beta }}{{\gamma _{SR}^2}}} \right]}^{ - \frac{1}{2}}}} \right)} \right] \cap {P_{SR,EH}} > \delta {P_{DR,EH}}} \right\}\\
\hspace{-2mm} &\approx \frac{{\left( {1 - {P_{ES}}} \right)}}{2}\left[ {\sum\limits_{m = 1}^M {{\omega _m}erfc\left( {\frac{{{W_1}{e^{{\kappa_m}}}}}{{\sqrt {8{W_1}{e^{{\kappa_m}}} + 8\beta } }}} \right)\frac{{\exp \left( { - \frac{{{e^{{\kappa_m}}}}}{{{{\overline P }_{SR,EH}}}} + {\kappa_m}{L_{sr}}} \right)}}{{\Gamma \left( {{L_{sr}}} \right)\overline P _{SR,EH}^{{L_{sr}}}}}\exp (\kappa_m^2) } } \right.\\
\hspace{-2mm}&\left. { -\hspace{-1mm} \sum\limits_{m = 1}^M {{\omega _m}}\hspace{-2mm} \sum\limits_{l = 0}^{{L_{rd}} - 1} \hspace{-2mm}{\frac{1}{{l!{{\left( {\delta {{\overline P }_{DR,EH}}} \right)}^{l}}}}er\!f\!c\left( \! {\frac{{{W_1}{e^{{\kappa_m}}}}}{{\sqrt {8{W_1}{e^{{\kappa_m}}} \!+\! 8\beta } }}} \!\right)\hspace{-1mm}\frac{{\exp\hspace{-1mm} \left(\hspace{-1mm} { - \frac{{{{\overline P }_{SR,EH}} + \delta {{\overline P }_{DR,EH}}}}{{\delta {{\overline P }_{SR,EH}}{{\overline P }_{DR,EH}}}}{e^{{\kappa_m}}} \!+\! {\kappa_m}({L_{sr}} \hspace{-1mm}+\! l)} \right)}}{{\Gamma \left( {{L_{sr}}} \right)\overline P _{SR,EH}^{{L_{sr}}}}}\exp (\kappa_m^2) \!+\! {O_M}} } \right].
 \label{eq:49func}
\end{aligned}
\end{equation}

Similarly, one can obtain the closed-form expression of $P_{e,SR}^{''}$ and $P_{e,RD}^{''}$ given in (\ref{eq:20func}) and (\ref{eq:21func}) by using the above method and $f_{\gamma_{RD}}(z)$ is derived in Appendix C.
In addition, though some simple approximated processes and using the above method, one can obtain the upper bound of $K_2$ as
\begin{equation}
\begin{aligned}
&{K_2} = E\left\{ {\frac{1}{2}er\!f\!c\left( {{{\left[ {\frac{8}{{{\gamma _{RD}}}} + \frac{{8\beta }}{{\gamma _{RD}^2}}} \right]}^{ - \frac{1}{2}}}} \right) \cap {P_{DR,EH}} > {P_{SR,EH}}/\delta } \right\}\\
& \hspace{-1mm}\le {O_M} + \sum\limits_{m = 1}^M {\sum\limits_{l = 0}^{^{{L_{rd}} - 1}} {\frac{{{\omega _m}{8^{{L_{rd}} - l}}\exp (\kappa _m^2)}}{{l!{\delta ^l}\Gamma \left( {{L_{sr}}} \right)\overline P _{SR,EH}^{{L_{sr}}}\overline P _{DR,EH}^l}}} }
\times \hspace{-1mm} \frac{{\exp\hspace{-1mm}  \left( \hspace{-1mm} { -\hspace{-1mm}  \left( {\frac{1}{{{{\overline P }_{SR,EH}}}} \hspace{-1mm} +\hspace{-1mm}  \frac{{{W_2}\left( {{e^{{\kappa _m}}} - {P_I}} \right)}}{{8\delta }} \hspace{-1mm} +\hspace{-1mm}  \frac{1}{{\delta {{\overline P }_{DR,EH}}}}} \right)\hspace{-1mm} {e^{{\kappa _m}}} \hspace{-1mm} +\hspace{-1mm}  {\kappa _m}\hspace{-1mm} \left( {{L_{sr}}\hspace{-1mm} + l} \right)} \right)}}{{{{\left( {{W_2}\left( {{e^{{\kappa _m}}} - {P_I}} \right){{\overline P }_{DR,EH}} + 8} \right)}^{{L_{rd}} - l}}}},
\label{eq:50func}
\end{aligned}
\end{equation}
where $W_2 =\frac{2}{\eta \theta P_{D}N_{0,rd}}$.

Furthermore, for the denominator in Eq.~(\ref{eq:46func}), one can obtain the probability of selecting S$\rightarrow$R link according to [43, eq. (3.351.2)], given by
\begin{equation}
\begin{aligned}
&\Pr ({P_{SR,EH}} > \delta {P_{DR,EH}})
 = \int_0^\infty  {\int_{\delta \zeta}^\infty  {{f_{{P_{SR,EH}}}}} } \left( v \right){f_{{P_{DR,EH}}}}\left(\zeta \right)dvd\zeta\\
 &= \frac{1}{{\Gamma \left( {{L_{sr}}} \right)\Gamma \left( {{L_{rd}}} \right)\overline P _{SR,EH}^{{L_{sr}}}\overline P _{DR,EH}^{{L_{rd}}}}}\sum\limits_{l = 0}^{{L_{sr}} - 1} {\frac{{\Gamma \left( {{L_{sr}}} \right)}}{{l!}}\frac{{{{\delta  }^{l}}}}{{{{\left( {\frac{1}{{{{\overline P }_{SR,EH}}}}} \right)}^{{L_{sr}} - l}}}}} \int_0^\infty  {{\zeta^{{L_{rd}} + l - 1}}} {e^{ - \frac{{\delta \zeta}}{{{{\overline P }_{SR,EH}}}}}}\exp\! \left( { - \frac{\zeta}{{{{\overline P }_{DR,EH}}}}} \right)\!d\zeta\\
 &= \frac{{\overline P _{SR,EH}^{{L_{rd}}}}}{{\Gamma \left( {{L_{rd}}} \right)}}\sum\limits_{l = 0}^{{L_{sr}} - 1} {\frac{{{\delta ^{l}}\overline P _{DR,EH}^{l}\Gamma \left( {{L_{rd}} + l} \right)}}{{l!{{\left( {\delta {{\overline P }_{DR,EH}} + {{\overline P }_{SR,EH}}} \right)}^{l + {L_{rd}}}}}}},
\label{eq:51func}
\end{aligned}
\end{equation}
The probability of selecting R$\rightarrow$D link is given by $P_{RD}=1-P_{SR}$.
Finally, through the combination of (\ref{eq:45func})-(\ref{eq:51func}), the end-to-end BER of two protocols can be obtained in Eqs.~(\ref{eq:24func}) and (\ref{eq:25func}).

\section{Derivation of $f_{\gamma_{RD}}(z)$}
In this appendix, the PDF expression of $\gamma_{RD}$ is derived. Here, the following formulas should be used [42, eq. (7.813)] and [43, eq. (4.257)]
\begin{equation}
\begin{aligned}
\int_0^\infty & {{u^{ - \rho }}{e^{ - B u}}\mathop G\nolimits_{\chi_1,\chi_2}^{\psi _1,\psi _2} \left( {\left. {C u} \right|_{{d_1},...,{d_{\psi _1}},...,{d_{\chi_2}}}^{{a_1},...,{a_{\psi _2}},...,{a_{\chi_1}}}} \right)du}
 = {B ^{\rho  - 1}}\mathop G\nolimits_{\chi_1 + 1,\chi_2}^{\psi_1,\psi _2 + 1} \left( {\left. {\frac{C }{B }} \right|\begin{array}{*{20}{c}}
{\rho ,{a_1},.....,{a_{\chi_1}}}\\
{{d_1},.....,{d_{\chi_2}}}
\end{array}} \right),
\label{eq:52func}
\end{aligned}
\end{equation}
and
\begin{equation}
\mathop G\nolimits_{\chi_1,\chi_2}^{\psi _1,\psi _2} \left( {{u^{ - 1}}|_{{d_s}}^{{a_r}}} \right) = \mathop G\nolimits_{\chi_2,\chi_1}^{\psi _2,\psi _1} \left( {u|_{1 - {a_r}}^{1 - {d_s}}} \right),
\label{eq:53func}
\end{equation}
where $\chi_1+\chi_2< 2\left ( \psi _1+\psi _2 \right )$, $|\arg(C) |< \left ( \psi _1+\psi _2-\frac{1}{2}\chi_1- \frac{1}{2}\chi_2\right )\pi $, $|\arg(B) |< \frac{1}{2}\pi $, $\Re\left ( d_{\Im}-\rho  \right )> -1$, and $G\left ( . \right )$ is the Meijer G-function defined in [42, eq. (9.301)].

The PDF of the transmitted power at R is computed as
\begin{equation}
{f_{{P_R}}}\left( x \right) = \frac{{{x^{L_{sr} - 1}}}}{{\Gamma \left( L_{sr} \right)\overline P _R^{L_{sr}}}}\exp \left( { - \frac{x}{{{{\overline P }_R}}}} \right),
\label{eq:54func}
\end{equation}
where $\overline{P}_{R}=\frac{\eta \theta P_{s}\Omega _{sr, l}}{d_{sr}^{\alpha }}-P_{I}$ and $\Omega _{sr,l}=E\left \{ \sum_{l=1}^{L_{sr}} h_{sr}^{2}\right \}$.

Let  $A=2/d_{rd}^{\alpha }$, $X=P_R$ and $Y=\sum_{l=1}^{L_{rd}}\   h_{rd,l}^{2}/N_{0,rd}$, then the PDF of the $Y$ is calculated as
\begin{equation}
{f_Y}\left( y \right) = \frac{{N_{0,rd}^{{L_{rd}}}{y^{{L_{rd}} - 1}}}}{{\Gamma \left( {{L_{rd}}} \right)\Omega _{rd,l }^{{L_{rd}}}}}\exp \left( { - \frac{{{N_{0,rd}}y}}{{{\Omega _{rd,l }}}}} \right),
\label{eq:55func}
\end{equation}
where $\Omega _{rd,l}=E\left \{ \sum_{l=1}^{L_{rd}} h_{rd}^{2}\right \}$.

Combining (\ref{eq:54func}) with (\ref{eq:55func}), the PDF of $\gamma_{RD}$ can be written as
\begin{equation}
{f_{{\gamma _{RD}}}}\left( z \right) = \frac{1}{A}\int_0^\infty  {\frac{1}{y}} {f_Y}\left( y \right){f_X}\left( {\frac{z}{{Ay}}} \right)dy.
\label{eq:56func}
\end{equation}

With the aid of (\ref{eq:52func}) and (\ref{eq:53func}), the closed-form expression of $f_{\gamma_{RD}}$ is expressed as
\begin{equation}
\begin{aligned}
{f_{{\gamma _{RD}}}}\left( z \right){\rm{ = }}&\frac{{N_{0,rd}^{{L_{sr}}}{z^{{L_{sr}} - 1}}}}{{{A^{{L_{sr}}}}\Gamma \left( {{L_{sr}}} \right)\Gamma \left( {{L_{rd}}} \right)\Omega _{rd,l }^{{L_{sr}}}\overline P _R^{{L_{sr}}}}}
\mathop G\nolimits_{0,2}^{2,0} \left( {\left. {\frac{{{N_{0,rd}}z}}{{A{{\overline P }_R}{\Omega _{rd,l }}}}} \right|\begin{array}{*{20}{c}}
 - \\
{{L_{rd}} - {L_{sr}},0}
\end{array}} \right).
\end{aligned}
\label{eq:57func}
\end{equation}

\end{appendices}


\begin{thebibliography}{1}
\bibliographystyle{IEEEbib}
\bibitem{8782883}
G.~{Cai}, Y.~{Fang}, J.~{Wen}, G.~{Han}, and X.~{Yang}, ``{QoS-aware
  buffer-aided relaying implant WBAN for healthcare IoT: Opportunities and
  challenges},'' \emph{IEEE Netw.}, vol.~33, no.~4, pp. 96--103, Jul. 2019.

\bibitem{1362898}
J.~N. {Laneman}, D.~N.~C. {Tse}, and G.~W. {Wornell}, ``{Cooperative diversity
  in wireless networks: Efficient protocols and outage behavior},'' \emph{IEEE
  Trans. Inf. Theory}, vol.~50, no.~12, pp. 3062--3080, Dec. 2004.

\bibitem{7314985}
Q.~F. {Zhou}, W.~H. {Mow}, S.~{Zhang}, and D.~{Toumpakaris}, ``{Two-way
  decode-and-forward for low-complexity wireless relaying: Selective forwarding
  versus one-bit soft forwarding},'' \emph{IEEE Trans. Wireless Commun.},
  vol.~15, no.~3, pp. 1866--1880, Mar. 2016.

\bibitem{8735871}
M.~{Atallah} and G.~{Kaddoum}, ``{Design and performance analysis of secure
  multicasting cooperative protocol for wireless sensor network
  applications},'' \emph{IEEE Wireless Commun. Lett.}, vol.~8, no.~5, pp.
  1468--1472, Oct. 2019.

\bibitem{7765086}
W.~{Wicke}, N.~{Zlatanov}, V.~{Jamali}, and R.~{Schober}, ``{Buffer-aided
  relaying with discrete transmission rates for the two-hop half-duplex relay
  network},'' \emph{IEEE Trans. Wireless Commun.}, vol.~16, no.~2, pp.
  967--981, Feb. 2017.

\bibitem{6408173}
N.~{Zlatanov} and R.~{Schober}, ``{Buffer-aided relaying with adaptive link
  selection-fixed and mixed rate transmission},'' \emph{IEEE Trans. Inf.
  Theory}, vol.~59, no.~5, pp. 2816--2840, May 2013.

\bibitem{8064682}
B.~{Kumar} and S.~{Prakriya}, ``{Performance of adaptive link selection with
  buffer-aided relays in underlay cognitive networks},'' \emph{IEEE Trans. Veh.
  Technol.}, vol.~67, no.~2, pp. 1492--1509, Feb. 2018.

\bibitem{6613630}
T.~{Islam}, A.~{Ikhlef}, R.~{Schober}, and V.~K. {Bhargava}, ``{Diversity and
  delay analysis of buffer-aided BICM-OFDM relaying},'' \emph{IEEE Trans.
  Wireless Commun.}, vol.~12, no.~11, pp. 5506--5519, Nov. 2013.

\bibitem{7317590}
T.~{Islam}, D.~S. {Michalopoulos}, R.~{Schober}, and V.~K. {Bhargava},
  ``{Buffer-aided relaying with outdated CSI},'' \emph{IEEE Trans. Wireless
  Commun.}, vol.~15, no.~3, pp. 1979--1997, Mar. 2016.

\bibitem{8328875}
N.~{Nomikos}, T.~{Charalambous}, D.~{Vouyioukas}, and G.~K. {Karagiannidis},
  ``{Low-complexity buffer-aided link selection with outdated CSI and feedback
  errors},'' \emph{IEEE Trans. Commun.}, vol.~66, no.~8, pp. 3694--3706, Aug.
  2018.

\bibitem{8390912}
L.~{Guntupalli}, M.~{Gidlund}, and F.~Y. {Li}, ``{An on-demand energy
  requesting scheme for wireless energy harvesting powered IoT networks},''
  \emph{IEEE Internet Things J.}, vol.~5, no.~4, pp. 2868--2879, Aug. 2018.

\bibitem{8628978}
M.~A. {Hossain}, R.~{Md Noor}, K.~A. {Yau}, I.~{Ahmedy}, and S.~S. {Anjum},
  ``{A survey on simultaneous wireless information and power transfer with
  cooperative relay and future challenges},'' \emph{IEEE Access}, vol.~7, pp.
  19\,166--19\,198, 2019.

\bibitem{7556958}
H.~{Liu}, K.~J. {Kim}, K.~S. {Kwak}, and H.~{Vincent Poor}, ``{Power
  splitting-based SWIPT with decode-and-forward full-duplex relaying},''
  \emph{IEEE Trans. Wireless Commun.}, vol.~15, no.~11, pp. 7561--7577, Nov.
  2016.

\bibitem{8463536}
X.~{Lan}, Q.~{Chen}, X.~{Tang}, and L.~{Cai}, ``{Achievable rate region of the
  buffer-aided two-way energy harvesting relay network},'' \emph{IEEE Trans.
  Veh. Technol.}, vol.~67, no.~11, pp. 11\,127--11\,142, Nov. 2018.

\bibitem{8107546}
R.~{Morsi}, D.~S. {Michalopoulos}, and R.~{Schober}, ``{Performance analysis of
  near-optimal energy buffer aided wireless powered communication},''
  \emph{IEEE Trans. Wireless Commun.}, vol.~17, no.~2, pp. 863--881, Feb. 2018.

\bibitem{8060568}
Z.~{Chen}, L.~X. {Cai}, Y.~{Cheng}, and H.~{Shan}, ``{Sustainable cooperative
  communication in wireless powered networks with energy harvesting relay},''
  \emph{IEEE Trans. Wireless Commun.}, vol.~16, no.~12, pp. 8175--8189, Dec.
  2017.

\bibitem{8122039}
Y.~{Feng}, V.~C.~M. {Leung}, and F.~{Ji}, ``{Performance study for SWIPT
  cooperative communication systems in shadowed Nakagami fading channels},''
  \emph{IEEE Trans. Wireless Commun.}, vol.~17, no.~2, pp. 1199--1211, Feb.
  2018.

\bibitem{6489506}
R.~{Zhang} and C.~K. {Ho}, ``{MIMO broadcasting for simultaneous wireless
  information and power transfer},'' \emph{IEEE Trans. Wireless Commun.},
  vol.~12, no.~5, pp. 1989--2001, May 2013.

\bibitem{8664088}
G.~{Shabbir}, J.~{Ahmad}, W.~{Raza}, Y.~{Amin}, A.~{Akram}, J.~{Loo}, and
  H.~{Tenhunen}, ``{Buffer-aided successive relay selection scheme for energy
  harvesting IoT networks},'' \emph{IEEE Access}, vol.~7, pp. 36\,246--36\,258,
  2019.

\bibitem{8226786}
Y.~{Liu}, Q.~{Chen}, and X.~{Tang}, ``{Adaptive buffer-aided wireless powered
  relay communication with energy storage},'' \emph{IEEE Trans. Green Commun.
  Netw.}, vol.~2, no.~2, pp. 432--445, Jun. 2018.

\bibitem{7008567}
Q.~{Yu}, D.~{Zhang}, H.~{Chen}, and W.~{Meng}, ``{Physical-layer network coding
  systems with MFSK modulation},'' \emph{IEEE Trans. Veh. Technol.}, vol.~65,
  no.~1, pp. 204--213, Jan. 2016.

\bibitem{5896023}
W.~{Guan} and K.~J.~R. {Liu}, ``{Performance analysis of two-way relaying with
  non-coherent differential modulation},'' \emph{IEEE Trans. Wireless Commun.},
  vol.~10, no.~6, pp. 2004--2014, Jun. 2011.

\bibitem{8036271}
G.~{Cai}, Y.~{Fang}, G.~{Han}, J.~{Xu}, and G.~{Chen}, ``{Design and analysis
  of relay-selection strategies for two-way relay network-coded DCSK
  systems},'' \emph{IEEE Trans. Veh. Technol.}, vol.~67, no.~2, pp. 1258--1271,
  Feb. 2018.

\bibitem{6482240}
P.~{Chen}, L.~{Wang}, and F.~C.~M. {Lau}, ``{One analog STBC-DCSK transmission
  scheme not requiring channel state information},'' \emph{IEEE Trans. Circuits
  Syst. I, Reg. Papers}, vol.~60, no.~4, pp. 1027--1037, Apr. 2013.

\bibitem{1362943}
{Yongxiang Xia}, C.~K. {Tse}, and F.~C.~M. {Lau}, ``{Performance of
  differential chaos-shift-keying digital communication systems over a
  multipath fading channel with delay spread},'' \emph{IEEE Trans. Circuits
  Syst. II, Exp. Briefs}, vol.~51, no.~12, pp. 680--684, Dec. 2004.

\bibitem{7426400}
F.~J. {Escribano}, G.~{Kaddoum}, A.~{Wagemakers}, and P.~{Giard}, ``{Design of
  a new differential chaos-shift-keying system for continuous mobility},''
  \emph{IEEE Trans. Commun.}, vol.~64, no.~5, pp. 2066--2078, May 2016.

\bibitem{5629387}
W.~{Xu}, L.~{Wang}, and G.~{Chen}, ``{Performance of DCSK cooperative
  communication systems over multipath fading channels},'' \emph{IEEE Trans.
  Circuits Syst. I, Reg. Papers}, vol.~58, no.~1, pp. 196--204, Jan. 2011.

\bibitem{7442517}
Y.~{Fang}, G.~{Han}, P.~{Chen}, F.~C.~M. {Lau}, G.~{Chen}, and L.~{Wang}, ``{A
  survey on DCSK-based communication systems and their application to UWB
  scenarios},'' \emph{IEEE Commun. Surveys Tuts.}, vol.~18, no.~3, pp.
  1804--1837, 3th Quarter 2016.

\bibitem{8606201}
H.~{Ma}, G.~{Cai}, Y.~{Fang}, J.~{Wen}, P.~{Chen}, and S.~{Akhtar}, ``{A new
  enhanced energy-detector-based FM-DCSK UWB system for tactile internet},''
  \emph{IEEE Trans. Ind. Informat.}, vol.~15, no.~5, pp. 3028--3039, May 2019.

\bibitem{8703732}
G.~{Cai}, Y.~{Fang}, J.~{Wen}, S.~{Mumtaz}, Y.~{Song}, and V.~{Frascolla},
  ``{Multi-carrier $M$-ary DCSK system with code index modulation: An efficient
  solution for chaotic communications},'' \emph{IEEE J. Sel. Topics Signal
  Process.}, vol.~13, no.~6, pp. 1375--1386, Oct. 2019.

\bibitem{8110728}
M.~{Herceg}, G.~{Kaddoum}, D.~{Vranješ}, and E.~{Soujeri}, ``{Permutation
  index DCSK modulation technique for secure multiuser high-data-rate
  communication systems},'' \emph{IEEE Trans. Veh. Technol.}, vol.~67, no.~4,
  pp. 2997--3011, Apr. 2018.

\bibitem{8468068}
M.~{Miao}, L.~{Wang}, M.~{Katz}, and W.~{Xu}, ``{Hybrid modulation scheme
  combining PPM with differential chaos shift keying modulation},'' \emph{IEEE
  Wireless Commun. Lett.}, vol.~8, no.~2, pp. 340--343, Apr. 2019.

\bibitem{8618388}
G.~{Cheng}, W.~{Xu}, C.~{Chen}, and L.~{Wang}, ``{SWIPT schemes for carrier
  index differential chaos shift keying modulation: A new look at the inactive
  carriers},'' \emph{IEEE Trans. Veh. Technol.}, vol.~68, no.~3, pp.
  2557--2570, Mar. 2019.

\bibitem{7973179}
W.~{Xu}, T.~{Huang}, and L.~{Wang}, ``{Code-shifted differential chaos shift
  keying with code index modulation for high data rate transmission},''
  \emph{IEEE Trans. Commun.}, vol.~65, no.~10, pp. 4285--4294, Oct. 2017.

\bibitem{7604059}
G.~{Kaddoum}, H.~{Tran}, L.~{Kong}, and M.~{Atallah}, ``{Design of simultaneous
  wireless information and power transfer scheme for short reference DCSK
  communication systems},'' \emph{IEEE Trans. Commun.}, vol.~65, no.~1, pp.
  431--443, Jan. 2017.

\bibitem{7109922}
L.~{Wang}, G.~{Cai}, and G.~R. {Chen}, ``{Design and performance analysis of a
  new multiresolution M-Ary differential chaos shift keying communication
  system},'' \emph{IEEE Trans. Wireless Commun.}, vol.~14, no.~9, pp.
  5197--5208, Sep. 2015.

\bibitem{6807959}
N.~{Zlatanov}, A.~{Ikhlef}, T.~{Islam}, and R.~{Schober}, ``{Buffer-aided
  cooperative communications: Opportunities and challenges},'' \emph{IEEE
  Commun. Mag.}, vol.~52, no.~4, pp. 146--153, Apr. 2014.

\bibitem{7365416}
N.~{Nomikos}, T.~{Charalambous}, I.~{Krikidis}, D.~N. {Skoutas},
  D.~{Vouyioukas}, M.~{Johansson}, and C.~{Skianis}, ``{A survey on
  buffer-aided relay selection},'' \emph{IEEE Commun. Surveys Tuts.}, vol.~18,
  no.~2, pp. 1073--1097, 2th Quarter 2016.

\bibitem{8830394}
X.~{Lan}, Q.~{Chen}, L.~{Cai}, and L.~{Fan}, ``{Buffer-aided adaptive wireless
  powered communication network with finite energy storage and data buffer},''
  \emph{IEEE Trans. Wireless Commun.}, vol.~18, no.~12, pp. 5764--5779, Dec.
  2019.

\bibitem{8718530}
D.~{Bapatla} and S.~{Prakriya}, ``{Performance of energy-buffer aided
  incremental relaying in cooperative networks},'' \emph{IEEE Trans. Wireless
  Commun.}, vol.~18, no.~7, pp. 3583--3598, Jul. 2019.

\bibitem{Papoulis1984Probability}
A.~Papoulis, \emph{{Probability, random variables, and stochastic processes.
  2nd ed.}}, 1984.

\bibitem{Abramowitz}
M.~Abramowitz, I.~A. Stegun, and J.~E. Romain, \emph{{Handbook of mathematical
  functions, with formulas, graphs, and mathematical tables}}.\hskip 1em plus
  0.5em minus 0.4em\relax New York: NY, USA: Dover, 1965.
\end{thebibliography}

\end{document}